\documentclass[preprint2]{aastex}

\def\FFr{\displaystyle \frac}

\begin{document}

\title{On the Physical Nature of the Source of Ultraluminous X-ray
Pulsations}

\author{G.Ter-Kazarian}\affil{Ambartsumian Byurakan
Astrophysical Observatory, Byurakan, 378433, Aragatsotn District,
Armenia\\ gago\_50@yahoo.com}

\begin{abstract}
To reconcile the observed unusual high luminosity of NuSTAR  X-ray
pulsations from M82X-2 with the most extreme violation of the
Eddington limit, and in view that the persistent X-ray radiation
from M82X-2 almost precludes the possibility of common pulsars, we
tackle the problem by the implications of {\em microscopic theory of
black hole} (MTBH). The preceding developments of MTBH are proved to
be quite fruitful for the physics of ultra-high energy (UHE)
cosmic-rays. Namely, replacing a central singularity by the
infrastructures inside event horizon, subject to certain rules, MTBH
explains the origin of ZeV-neutrinos which are of vital interest for
the source of UHE-particles. The M82X-2 is assumed to be a spinning
intermediate mass black hole resided in final stage of growth. As a
corollary, the thermal blackbody X-ray emission arisen due to the
rotational kinetic energy of black hole escapes from event horizon
through the vista to outside world that detected as ultraluminous
X-ray pulsations. The M82X-2 indeed releases $\sim 99.6\%$ of its
pulsed radiative energy predominantly in the X-ray bandpass $0.3-30$
keV. We derive a pulse profile and give a quantitative account of
energetics and orbital parameters of the semi-detached X-ray binary
containing a primary accretor M82X-2 of inferred mass $M\simeq
138.5-226\,M_{\odot}$ and secondary massive, $M_{2}> 48.3-
64.9\,M_{\odot}$, O/B-type donor star  with radius of $R> 22.1-
25.7\,R_{\odot}$, respectively. We compute the torque added to
M82X-2 per unit mass of accreted matter which yields the measured
spin-up rate.
\end{abstract}


\section{Introduction}
The physical nature of the accreting off-nuclear point sources in
nearby galaxies, so-called the ultraluminous X-ray sources (ULXs),
has been an enigma because of their high energy output characterized
by super-Eddington luminosities up to two orders of magnitude higher
than those observed from Galactic X-ray binaries, and unusual soft
X-ray spectra with blackbody emission around $\lesssim 0.3$ keV and
a downturn above $\sim 5$ keV ~(e.g.~\citet{FGT,King,SwGh,Rob,GRD,
FS,Li,WRMH,LBB,MPS}). In spite of significant efforts in more than
three decade since the discovery of ULXs, the astronomers have not
yet clarified their nature. Without care of the historical justice
and authenticity, it should be emphasized that the ULX sources have
attracted a great deal of observational and theoretical attention,
in part because their luminosities suggest that they may harbor
intermediate mass black holes (IMBHs) with an ubiquitous feature of
the mass fits of more than $10^{2} -
10^{4}\,M_{\odot}$~\citep{FS,CM,Mak}. A strong argument in favor of
IMBHs is the presence of a soft, $0.1–0.2$ keV component in their
spectra~\citep{KCPZ,MFMF,MFM}. The hyperluminous X-ray sources with
luminosities $\geq\, 10^{41}\, {\rm erg\, s^{-1}}$ are thought to be
amongst the strongest IMBH candidates~\citep{Mat}. Assuming the
emission is isotropic, in general, the extreme luminosities of ULXs
suggest either the presence of IMBHs and sub-Eddington accretion
~(e.g.~\citet{KSh,ColMil,Dew,SRW,
PPD,PBH,KYH,CPP,FK10,LBB,FS,WMHF,PSM}), or stellar-mass black holes
($M \leq 10\,M_{\odot}$)  that are either breaking  or circumventing
their Eddington limit via somewhat geometric beaming of accretion
flow~\citep{Beg,King,Oka,Rob,ZR,SRW,Pout}. The latter remains still
a relatively poorly understood regime.~\citet{SRWGS} present the
results from an X-ray and optical study of a new sample of eight
extreme luminosity ULX candidates, which were selected as the
brightest ULXs (with $L_{X}> 5\times 10^{40}\, {\rm erg \,s^{-1}}$)
located within $100$ Mpc identified in a cross correlation of the
2XMM-DR1 and RC3 catalogues. These objects seemed to be amongst the
most plausible candidates to host larger, IMBHs.  But, future
improved observations are needed yet to decide on the issue. The
quasi-periodic oscillations (QPOs) and frequency breaks in
XMM-Newton power-density spectra of ULXs with luminosity, $L_{X}
\geq  10^{40} \,{\rm erg\, s^{-1}}$, suggest that the black hole
masses are more consistent with IMBHs than stellar-mass black
holes~\citep{ColMil}. Or, using high sensitive dataset in the Fe K
region obtained in scope of a long Suzaku program on Holmberg IX X-1
for any luminous, isolated ULX to date, ~\citet{WMHF} find no
statistically significant narrow atomic features in either emission
or absorption across the 5–9 keV energy range. They conclude,
therefore, that the models of spherical super-Eddington accretion
can be rejected, as can wind-dominated spectral models. The lack of
iron emission implies that the stellar companion in sub-Eddington
accretion onto an IMBH is unlikely to be launching a strong wind,
and therefore the black hole must primarily accrete via roche-lobe
overflow. There are also several other new results of more recent
studies: e.g.~\citet{WGS} demonstrate that if the transferred mass
is efficiently converted to X-ray luminosity (with disregard of the
classical Eddington limit) or if the X-rays are focused into a
narrow beam then binaries can form extreme ULXs with the X-ray
luminosity of $L_{X} \gtrsim 10^{42}$ erg\, s$^{-1}$. They envisage
that these systems are not only limited to binaries with
stellar-origin black hole accretors, but also can be identified with
neutron star systems. For the latter the typical donors are evolved
low-mass $(2\,M_{\odot})$ helium stars with Roche lobe overflow rate
of $\sim 10^{-2}\,M_{\odot}$yr$^{-1}$. This study does not prove
that any particular extreme ULX is a regular binary system. Or,
~\citet{FLK} performed a population synthesis studies of ULXs with
neutron star accretors. Combining binary population synthesis and
detailed mass-transfer models, they conclude that the binary system
that formed M82X-2 is most likely less than $50$\,Myr old and
contains a donor star which had an initial mass of approximately
$8-10\, M_{\odot}$, while the NS's progenitor star had an initial
mass in the $8-25\, M_{\odot}$ range. Or,~\citet{MST} study
properties of luminous X-ray pulsars using a simplified model of the
accretion column. The resulting luminosity of NS pulsars may can
reach values of the order of $10^{40}$ erg s$^{-1}$ for the
magnetar-like magnetic field, since the equilibrium, where the
Alfv\'{e}n radius matches the co-rotation radius, indicates a
magnetic field of $B \gtrsim 10^{14} $G, and long spin periods ($P
\gtrsim 1.5 s$). They conclude that a substantial part of ULXs are
accreting neutron stars in binary systems. All these proposals have
own advantages and difficulties. Nevertheless, no single theory has
been invented yet which successfully addresses the solution to the
problems involved. The aforementioned studies are not exception to
the rule that as phenomenological approaches they suffer from own
difficulties. Namely, they are strongly model dependent, and subject
to many uncertainties and controversies. The physics is obscured by
multiple arbitrary assumptions and proliferation of a {\em priori}
free parameters involved, while a consistent complete theory would
not have so many free parameters. So, the issue is still
controversial, and these exotic results and conclusions are still
lame and for sure not proven by the authors. For
example,~\citet{WGS} have evoked the possibility that the
transferred mass is efficiently accreted onto a compact object and
converted to X-ray luminosity in the full range of possible mass
accretion rates, which is in contrast with the generally accepted
view that the conversion efficiency decreases with increasing mass
accretion rate (e.g., ~\citet{Pout}). Or, in case of model of the
accretion column, a necessary condition for the X-ray luminosity to
exceed the Eddington limit is a certain degree of asymmetry in the
distribution of matter over the Alfv\'{e}n surface~\citep{BS}, which
is a qualitative picture, perhaps, but not quantitative one, or a
$B\gtrsim 10^{14}$ G magnetic field of a magnetar-like pulsar is
rather above the quantum limit, etc. We will not be concerned here
with the actual discussion of these results, which require further
extensive and careful analysis, in order one could verify the
proposed estimates and constrain model parameters. There were still
many open key questions arisen inevitably that we have no
understanding of ULX physics. A main physical issue whether ULXs are
powered by IMBHs or normal stellar black holes to date is
unresolved, primarily because we do not have dynamical mass
measurements of the compact objects that power ULXs. So, it is
premature to draw conclusions and only time will tell whether any of
these intriguing proposals is correct and which of the hypothesized
ULX scenario is actually realized in nature.

\subsection{The ultraluminous X-ray pulsations}
The most striking is the recent revolutionary NuSTAR discovery of
the first rare and mighty ultraluminous X-ray pulsations with the
maximum luminosity $\widetilde{L}(3-30 {\rm keV})= 4.9\times
10^{39}\, {\rm erg\, s^{-1}}$, of average period $1.37$ s with a
$2.5$-day sinusoidal modulation~\citep{NuS}, coming from an
ultraluminous source, NuSTAR J095551+6940.8, located nearby
starburst galaxy M82 (NGC 3034). The pulsed emission centroid is
spatially consistent with the location of a variable M82X-2 which
further secures the association of the pulsating source with M82X-2.
Detection of coherent pulsations, a binary orbit, and spin-up
behavior indicative of an accretion torque unambiguously, allow to
feature the M82X-2 as mass-exchange binary that contains a
nondegenerate secondary donor star. Eventually, in addition to the
orbital modulation, about an evident linear spin-up of the pulsar
was reported by~\citep{NuS}, with $\dot{p}\simeq -2\times 10^{-10}
{\rm s/s}$ over the interval from modified Julian days 56696 to
56701 when the pulse detection is most significant. Phase connecting
the observations enables detection of a changing secular spin-up
rate over a longer timespan as well as erratic variations. Future
observations will show whether the current spin-up rate is secular.
~\citet{KLa} point out that the spin-up to luminosity ratio
$10^{-50}$ (erg · s)$^{-1}$ is an order of magnitude lower than the
typical ratio observed in the X-ray pulsars, which makes an
interpretation of the data in terms of a strongly magnetized X-ray
pulsar quite challenging. In absolute terms, this spin-up rate is
orders of magnitude higher than the values measured in the usual
accretion powered X-ray pulsars~\citep{BeA,Zi}.

\subsection{Key objections} A current understanding of NuSTAR discovery
is quite complicated. At first glance it seems as though the NuSTAR
team has demonstrated that the super-Eddington accretion is also
possible in ULXs hosting a neutron star, because it is generally
believed that the pulsating X-ray sources are magnetic neutron stars
which are accreting matter from a binary
companion~\citep{PR,DO,LPP,Lam}. Therefore, it seems that there is
nothing left but M82X-2, which until recently astronomers had
thought was a black hole, is the brightest magnetic neutron star
system ever recorded. This point of view is widely quoted in
literature and, at first sight, seems eminently reasonable. However,
deeper examination raises several disturbing questions, if the above
result is really valid. Even though, if for a moment we take the
classical accreting neutron star pulsar system as a basic
assumption, note that this model is not flawless. The actual
mechanism by which pulsars convert the rotational energy of the
neutron star into the observed pulses is poorly understood. Many
theoretical models have been proposed that account for such
features, but no single one is compelling~\citep{SW,Mich}.

Explaining periodic source M82X-2 that obviously has black hole
energetics with a $\sim 1.4\,M_{\odot}$ compact object using the
accreting neutron star pulsar systems is extremely challenging,
because of several problems we encountered. The difficulty becomes
apparent when one follows the three objections, which together
constitute a whole against the claim that M82X-2, perhaps, is a
common pulsar: 1) The NuSTAR team discovery is the most extreme
violation of the so-called Eddington limit, i.e. the pulsed
luminosity of M82X-2 reaches about $\sim 26.9$ times brighter than
the theoretical threshold at the spherical accretion for $\sim
1.4\,M_{\odot}$ stellar-mass black holes  where the outward pressure
from radiation balances the inward pull of gravity of the pulsar.
The accretion is inhibited once radiation force is equal or grater
than gravity force. 2) The difficulty is brought into sharper focus
by considering the association with M82X-2, which is featured with
high luminosity ($\simeq 1.8\times 10^{40}\, {\rm erg\, s^{-1}}$) of
additional persistent continuous broad X-ray radiation observed
earlier from its active state~\citep{NuS}. This more compelling
argument in somehow or other implies the luminosity $\sim 100$ times
if compared to the Eddington limit. Such a collimation $(\sim 100)$,
which is comparable to that obtained for black holes (e.g.,
~\citet{Oh,Sad}), would be needed to explain M82X-2 as beamed
radiation from neutron star. 3) Equally noteworthy that the centroid
of the persistent X-ray emission is between M82X-2 and M82X-1. If
M82X-1 is indeed harbors plausible IMBH,
(e.g.~\citet{PPD,PBH,Rob,CPP,FK10,FS,PSM}), we expect the similarity
of the persistent X-ray properties of the M82X-1 and M82X-2 to imply
that the non-pulsed emission from the latter originates in the
accretion disc, as it must in the black hole M82X-1. In this sense,
the NuSTAR discovery is unexpected and still hard to be explained in
the context of magnetic neutron star pulsar model. The fraction of
ULXs powered by neutron stars must be considered highly uncertain
and many details of this scenario remain poorly understood. Added to
this was the fact that NuSTAR finding may indeed not be rare in the
ULX population. In the future astronomers also will look at more
ULXs, and it is a matter of time until they could prove an expected
ubiquitous feature of even more energetic ULX pulsations, as being
common phenomena in the Universe. If confirmed, this would support a
scenario in which the more ULXs beat with the pulse of black holes
rather than magnetized neutron star systems. What if NuSTAR
detection might radically change one's view of Nature. With this
perspective in sight, it is wise in the case of M82X-2 to place
constraints on the likelihood of the magnetic neutron star pulsar
systems, so the model of common pulsar will be tested critically.

\subsection{Result}
Putting apart the discussion of inherent problems of the mass
scaling of the black holes in ULXs, which is beyond the scope of
this report, we approach the M82X-2 issue from the standpoint of
black holes rather than magnetic neutron star pulsar systems. To
reconcile the observed unusual high pulsed luminosity with the above
mentioned violation of the Eddington limit, we examine the physics
which is at work in ultraluminous pulsations by assuming that M82X-2
is being a spinning intermediate-mass black hole (SIMBH). If a black
hole of intermediate mass will be an exact law of Nature, it is
certainly an attractive scenario. Fore example, ~\citet{KYH} suggest
that the M82X-2 is a binary system with a black hole accretor.
Assuming the persistent emission is isotropic, the X-ray luminosity
$\simeq 10^{40}\, {\rm erg\, s^{-1}}$ implies that the compact
object is a $> 100 M_{\odot}$ IMBH in the low/hard state. However,
one may be tempted to argue truly that most of the issues and
objections raised above cannot be solved in the framework of
conventional black hole models, to which we refer as
phenomenological black hole models (PBHMs) (see subsect.2.1). The
coherent periodicity obviously rules out PBHM, because 1) black
holes do not radiate; 2) the spinning black holes are axisymmetric
and have no internal structure on which to attach a periodic
emitter. Orbital motion, whether modulating some emission mechanism
directly or exciting short-period pulsations, would decay very
quickly due to gravitational radiation. With this in mind, we
revisited the MTBH which completes PBHM by exploring the most
important processes of spontaneous breaking of gravitation gauge
symmetry and rearrangement of vacuum state at huge energies. One of
the purposes of this report is to motivate and justify the further
implications of MTBH framework to circumvent the alluded obstacles
without the need for significant breaking of Eddington limit. We
will proceed according to the following structure. To start with,
some latest developments on IMBHs are discussed in subsection 1.4.
We provide an analysis aimed at clarifying the current situation. To
make the rest of paper understandable, section 2 deals with a brief
review of key objectives of the MTBH framework, in particular, in
relevance to IMBH seeds as a guiding principle. In section 3 we set
out to examine the ultraluminous pulsations powered by M82X-2,
constituting mass-exchange binary with the O/B-type donor star. We
discuss a basic geometry which describes the rotating axisymmetric
black holes and bring some observations which comprise the whole of
the case. We derive a general profile of pulsed luminosity of
M82X-2, give a quantitative account of a potential dynamical mass
scaling of M82X-2 and other energetics, estimate the mass of
companion and the orbit parameters of the mass-exchange binary,
discuss the measured spin-up rate, and calculate the torque added to
M82X-2 per unit mass of accreted matter. Concluding remarks are
presented in section 4.

\subsection{Some latest developments on IMBHs}
Transient behavior of M82X-2 with a massive companion donor star
likely requires an IMBH, even though the physical interpretation of
the latter is still controversial. If IMBHs exist they have
important implications for the dynamics of stellar clusters and the
formation of supermassive black holes. These objects are of
particular interest because this population would fall within the
gap between the two principal black hole populations, i.e.
super-massive black holes (SMBHs), with masses of
$10^{6}-10^{10}\,M_{\odot}$,  and stellar mass black holes, with
masses of $\sim 10\,M_{\odot}$. A ROSAT/HRI catalog of
ULXs~\citep{CM} provides a clue to fill in the missing population of
the IMBHs. Currently there are very few well studied IMBHs, two
examples are NGC 4395~\citep{FH}  and POX 52~\citep{KSB,BHR}. The
IMBHs have non-standard spectra, which show a cutoff at a few
keV~\citep{SRW}. Note that other black holes, stellar as well as
super-massive, at a few per cent of Eddington luminosity have hard
power-law-like spectra~\citep{ZJP}. Some ULXs show a harder, $1–4$
keV thermal component with the corresponding radius of only 30–40
km~\citep{Mak,SRW}. It is not clear to date how IMBHs would form.
They have remained observationally elusive, with dynamical evidence
for such objects in large globular clusters still the subject of
some dispute. On the one hand, to produce an massive IMBH of
$10^{2}-10^{4}\,M_{\odot}$, the core collapse of an isolated star in
the current epoch is not a viable process, which is how the stellar
black holes are thought to form, because of the lack of very massive
stars. Their environments lack the extreme conditions, i.e. high
density and velocities observed at the centers of galaxies, which
seemingly lead to the formation of suppermassive black holes. There
are two conventional scenarios for the formation of IMBHs. The first
is the merging of stellar mass black holes and other compact objects
by means of gravitational radiation. The second one is the runaway
collision of massive stars in dense stellar clusters and the
collapse of the collision product into an IMBH. But, in fact, most
ULX host galaxies do not even have stellar clusters sufficiently
massive and compact to satisfy the requirements for runaway core
collapse. These objects could be formed inside clusters that have
since dispersed. However, the evaporation timescale of such clusters
would be too long to explain the observed association of many ULXs
with young $(\,\leq\, 20 \,{\rm Myr})$ stellar populations. For
example, one of the strongest candidates for IMBH to date is the
hyperluminous X-ray source HLX1 or 2XMM J011028.1-460421, possibly
located in the galaxy ESO243-49, or, at least, projected inside the
$\mu_{B} = 25.0$ mag arcsec$^{-2}$ surface brightness isophote of
that galaxy~\citep{Farr,Gode}. ~\citet{SHGK} report a discovery of
the likely optical counterpart to this source. In their analysis of
the UV emission in ESO243- 49, they determine the optical flux, and
the X-ray/optical flux ratio, and conclude that the X-ray source
belongs to ESO243-49, seemingly located inside a massive star
cluster. A metal-free population III stars formed in the very early
Universe could reach masses of a few hundred $M_{\odot}$, above the
pair-instability limit, and thus may have collapsed into
IMBHs~\citep{MR}. It was then expected that in young and dense star
clusters, dynamical friction could lead to massive stars sinking
towards the center and undergoing runaway collisions and mergers on
timescales $10^{6}\,{\rm yr}$. Dynamical evidence for IMBHs with
masses $\sim 10^{4}\,M_{\odot}$ has been proposed for a few globular
clusters, e.g., G1 in M31 at the $1.5\sigma$ significance
level~\citep{Geb}, although the issue is still controversial, and
these results require further extensive and careful investigation of
the stellar density and velocity distributions in globulars, being
of the highest importance in understanding of the formation of
IMBHs~\citep{AvM}. In massive and/or compact globular clusters, a
central seed black hole may grow by up to a factor of 100 via
accretion of gas lost by the first generation of cluster stars in
their red-giant phase~\citep{Vesp}; or, IMBHs may wander in the halo
of major galaxies, after tidal stripping of merging satellite dwarfs
that contained nuclear BHs ~\citep{KD,Bell}. Although this seems one
of the most plausible explanations for the above mentioned source
HLX-1, in general, currently we do not have a complete theoretical
interpretation of physical nature of growth of black hole seeds.
Therefore, the MTBH framework may become of eminent physical
significance in tackling this problem.

\section{MTBH, Revisited:\,the implications for
IMBHs}  The aim of the present section is to recount some of the
highlights behind the MTBH for the benefit of the reader, as a
guiding principle to make the rest of paper understandable. We shall
see how a IMBH seed is thought to form. There are several important
topics not touched upon here, which will eventually benefit from a
proposed gravitation theory. Although some key theoretical ideas
were introduced with a satisfactory substantiation, we have also
attempted to maintain a balance being overly detailed and overly
schematic.

\subsection{Phenomenological model of black holes}
From its historical development, up to current interests, the
efforts in the active galactic nuclei (AGN) physics have evoked the
study of a major unsolved problem of how efficiently such huge
energies observed can be generated. This energy scale severely
challenges conventional source models. The huge energy release from
compact regions of AGNs requires extremely high efficiency
(typically $\geq 10$ per cent) of conversion of rest mass to other
forms of energy. This led early on to suggestions that AGNs are
powered by SMBHs, with masses of $10^{6}-10^{10}\,M_{\odot}$. A
complex study of AGN evolution requires a comprehensive
understanding of a new field of astrophysics, so-called black hole
demography. This studies the important phenomena of the birth and
growth of black holes, the merging of stellar mass black holes and
their evolutionary connection to other objects in the Universe.
These ideas gather support especially from a breakthrough made in
recent observational and computational efforts on understanding of
coevolution of black holes and their host galaxies, particularly
through self-regulated growth and feedback from accretion-powered
outflows (e.g.~\citet{NT,Ves,VLN,VN,Vol1,Sha,Kel,Na,TM,Wil1,DV}. In
these models, at early times the properties of the assembling SMBH
seeds are more tightly coupled to properties of the dark matter halo
as their growth is driven by the merger history of halos.

General relativity (GR) has stood the test of time and can claim
remarkable success, although there are serious problems concerning
the role of singularities or black holes. This state of affairs has
not much changed up to present and proposed abundant models are not
conductive to provide non-artificial and unique recipe for resolving
controversial problems of energy-momentum conservation laws of
gravitational interacting fields, localization of energy of
gravitation waves and also severe problems involved in quantum
gravity. A tacit assumption of theoretical interpretation of
aforementioned astrophysical scenarios is a general belief
(e.g.~\citet{Rees,Shap,MuBi,AlNa} and references therein) reinforced
by statements in textbooks, that a longstanding standard PBHM  can
describe the growth of accreting black hole seeds. In the framework
of GR, the PBHM implies the most general Kerr-Newman black hole
model, with the integral parameters of total mass ($M$), angular
momentum ($J$) and charge ($Q$), still has to put in by hand. But
such beliefs are suspect and should be critically re-examined. Even
though being among the most significant advances in astrophysics, it
is rather surprising that PBHM is routinely used to explore this
problem as though it cannot be accepted as convincing model for
addressing the black hole growth, because in this framework the very
source of gravitational field of the black hole is a kind of
meaningless curvature singularity at the center of the stationary
black hole, which is hidden behind the event horizon. The theory
breaks down inside the event horizon which is causally disconnected
from the exterior world. Either the Kruskal continuation of the
Schwarzschild ($J=0,\,Q=0$) metric, or the Kerr ($Q=0$) metric, or
the Reissner-Nordstrom ($J=0$) metric, show that the static
observers fail to exist inside the horizon. Any object that
collapses to form a black hole will go on to collapse to a
singularity inside the black hole. Any timelike worldline must
strike the central singularity which wholly absorbs the infalling
matter. Therefore, the ultimate fate of collapsing matter once it
has crossed the black hole surface is unknown. So, one should
deliberately forebear from presumption of exotic hypothetical growth
behavior of black holes, which seems nowhere near true if one
applies the phenomenological model. This ultimately disables any
accumulation of matter in the central part and, thus, neither the
growth of black holes nor the increase of their mass-energy density
could occur at accretion of outside matter, or by means of merger
processes. Yet it is still thought provoking how can one be sure
that some hitherto unknown source of pressure does not become
important at huge energies and halt the infinite collapse. That in
PBHM  there is no provision for growth behavior of black holes, is
because in it one assigned only an insufficient attributes to this.
The PBHM is a rather restricted model and one needs to realise that
if one can gain insight into exploring a new process of spontaneous
breaking of gravitation at huge energies, one has then made room for
growth and merging behavior of black holes. To fill the void which
the standard PBHM presents, thus, one plausible idea to innovate the
solution to alluded key problems would appear to be such breaking
mechanism, and thereof for that of rearrangement of vacuum state
itself.

\subsection{Proposed gravitation theory}
In this subsection we present a brief outline of the key ideas of
the underlying theory, which yields the gravitational interaction at
huge energies drastically different from earlier suggested schemes.
This theory involves a drastic revision of a role of local internal
symmetries in physical concept of curved geometry, and explores the
most important processes of spontaneous breaking of gravitation
gauge symmetry and rearrangement of vacuum state. It was originally
proposed by~(\citet{b251,b50,b502} and references therein), and
thoroughly discussed in ~\citep{gago2,gago3}. This extension,
suitable for applications in extremely high energy astrophysics, is
a bold assumption in its own right. In its present formulation, this
theory exploits the language of the fundamental geometric
structure\,-\,{\em distortion gauge induced fiber-bundle}, which
provides a modified gravitational theory, as a corollary of the
spacetime deformation/distortion framework, suggested
by~\citet{gago3} (see also~\citet{gago4,gago5}). It should be
emphasized, that the standard Riemannian (and its extensions) space
interacting quantum field theory cannot be a satisfactory ground for
addressing the problems in quest. The difficulties associated with
this step are notorious, however, these difficulties are technical.
In the main, they stem from the fact that Riemannian geometry, in
general, does not admit a group of isometries, and that it is
impossible to define energy-momentum as Noether local currents
related to exact symmetries. This, in turn, posed severe problem of
non-uniqueness of the physical vacuum and the associated Fock space.
A definition of positive frequency modes cannot, in general, be
unambiguously fixed in the past and future, which leads to
$|in>\neq|out>,$ because the state $|in>$ is unstable against decay
into many particle $|out>$ states due to interaction processes
allowed by lack of Poincar\'{e} invariance. A non-trivial Bogolubov
transformation between past and future positive frequency modes
implies that particles are created from the vacuum and this is one
of the reasons for $|in>\neq|out>$. Keeping in mind aforesaid, we
develop on the framework of the {\em general gauge principle}, as
{\em from first principles}:\, We consider the principal fiber
bundle with the semi-Riemannian spacetime, $V_{4}$, as the base
space, and with the structure group, $G_{V}$, generated by the
hidden local internal symmetry of two-parameter abelian local group
$U^{loc}(2)=U(1)_{Y}\times \overline{U}(1)\equiv U(1)_{Y}\times
diag[SU(2)]$. This group is implemented on the flat space, $M_{4}$.
The collection of matter fields of arbitrary spins take values in
standard fiber. To involve a drastic revision of the role of gauge
fields in the physical concept of the spacetime
deformation/distortion, we generalize the standard gauge scheme via
the concept of distortion gauge field which acts on the external
spacetime groups. The group $U^{loc}(2)$ has two generators, the
third component $T^{3}$ of isospin $\vec{T}$ related to the Pauli
spin matrix $\frac{\vec{\tau}}{2}$, and hypercharge $Y$ implying $
Q^{d}=T^{3}+\frac{Y}{2}, $ where $Q^{d}$ is the {\em distortion
charge} operator assigning the number -1 to particles, but +1 to
anti-particles. We connect the structure group $G_{V}$, further, to
the nonlinear realization of the Lie group $G_{D}$ of {\em
distortion} of the spacetime~(e.g.\citet{gago2,gago3}), i.e. we
extend the curvature of the spacetime continuum to general {\em
distortion} as the theory of spontaneous breaking of {\em
distortion} symmetry. This can be achieved by non-linear
realizations of the {\em distortion} group $G_{D}$. The nonlinear
realization technique or the method of phenomenological
Lagrangians~\citep{rf55,rf66,Sal,Ish,O,V,rf77} provides a way to
determine the transformation properties of fields defined on the
quotient space. In accord, we treat the distortion group $G_{D}$ and
its stationary subgroup $H=SO(3)$, respectively, as the dynamical
group and its algebraic subgroup. The fundamental field is
distortion gauge field and, thus, all the fundamental gravitational
structures in fact - the metric as much as the coframes and
connections - acquire a {\em distortion-gauge induced} theoretical
interpretation. We study the geometrical structure of the space of
parameters in terms of Cartan's calculus of exterior forms and
derive the Maurer-Cartan structure equations. There is a
reciprocity, treated in the Maurer-Cartan nonlinear structure
equations, between the formal mathematical structure of distortion
fields on the one hand, and the Goldstone fields, on the other. The
group $U^{loc}(2)$  entails two neutral gauge bosons of
$\overline{U}(1)$, or that coupled to $T^{3}$, and of $U(1)_{Y}$, or
that coupled to the hypercharge $Y$. Spontaneous symmetry breaking
can be achieved by introducing the neutral complex scalar Higgs
field. Minimization of the vacuum energy fixes the non-vanishing
vacuum expectation value, which spontaneously breaks the theory,
leaving the $U(1)_{d}$ subgroup intact, i.e. leaving one Goldstone
boson. Consequently, the left Goldstone boson is gauged away from
the scalar sector, but it essentially reappears in the gauge sector
providing the longitudinally polarized spin state of one of gauge
bosons that acquires mass through its coupling to Higgs scalar.
Thus, the two neutral gauge bosons were mixed to form two physical
orthogonal states of the massless component of {\em distortion}
field, $(a)$ ($M_{a}= 0$), which is responsible for gravitational
interactions, and its massive component, $(\bar{a})$
($M_{\bar{a}}\neq 0$), which is responsible for the, so-called {\em
inner distortion} (ID)-regime. Hence, a substantial change of the
properties of the spacetime continuum besides the curvature may
arise at huge energies. This theory is renormalizable, because gauge
invariance gives conservation of charge, also ensures the
cancelation of quantum corrections that would otherwise result in
infinitely large amplitudes. Without careful thought we expect that
in this framework the renormalizability of the theory will not be
spoiled in curved space-time as well, because, the infinities arise
from ultra-violet properties of Feynman integrals in momentum space
which, in coordinate space, are short distance properties, and
locally (over short distances) all the curved spaces look like {\em
maximally symmetric} (flat) space.

For brevity, we will not be further concerned  with the actual
details of this comprehensive theoretical framework, but only use it
as a backdrop to validate the theory with more observational tests.
For details, the interested reader is invited to consult the
original papers. Discussed gravitation theory is consistent with GR
up to the limit of neutron stars. But MTBH manifests its virtues
applied to the physics at huge energies. Whereas a significant
change of properties of spacetime continuum, ID-regime (at
$\bar{a}\neq 0$), arises simultaneously with the strong gravity
$(a\neq 0)$. The Compton length of the ID-field $\bar{a}(r)$ is
$\lambda_{\bar{a}}=\frac{\hbar}{m_{\bar{a}} c}\simeq 0.4 fm$.

\subsection{Crossing event horizon from inside the black hole}
Let us to discuss  in more details some principle issues in use.
Recall that according to GR, the event horizon is impenetrable
barrier for crossing from inside the black hole because of a
singularity arisen at the Schwarzschild radius (radius of a
non-rotating black hole) $R_{g}=2GM/c^{2}=2.95\times
10^{5}\,M/M_{\odot}$ cm. But in the framework of MTBH, this barrier
disappears when a matter found in the ID-region of space-time
continuum has undergone phase transition of II-kind and, thus, it
transformed to proto-matter. To obtain some feeling about this
phenomena, below we give more detailed explanation. In the framework
of MTBH, we are led to construct a formalism of unitary mapping of
the fields and their dynamics from $M_{4}$ to $V_{4}$, and vice
versa. The field equations follow at once from the total gauge
invariant Lagrangian in terms of Euler-Lagrange variations,
respectively on both $V_{4}$ and $M_{4}$ spaces~\citep{gago2,gago3}.
The Lagrangian of distortion gauge field defined on the flat space
is undegenerated Killing form on the Lie algebra of the group
$U^{loc}(2)$ in adjoint representation, which yields the equation of
distortion field. In the reminder of this subsection the quantities
referred to $V_{4}$ are denoted by wiggles, and left without wiggles
if they correspond to $M_{4}$. We are interested in the case of a
spherical-symmetric gravitational field $a_{0}(r)$ in presence of
one-dimensional space-like ID-field $\bar{a}$. In the case at hand,
one has the group of motions $SO(3)$ with 2D space-like orbits
$S^{2}$ where the standard coordinates are $\widetilde{\theta}$ and
$\widetilde{\varphi}$. The stationary subgroup of $SO(3)$ acts
isotropically upon the tangent space at the point of sphere $S^{2}$
of radius $\widetilde{r}$. So, the bundle $p:V_{4} \rightarrow
\widetilde{R}^{2}$ has the fiber $S^{2}=p^{-1}(\widetilde{x})$,
$\quad \widetilde{x}\in V_{4}$ with a trivial connection on it,
where $\widetilde{R}^{2}$ is the quotient-space $V_{4}/SO(3)$.
Considering the equilibrium configurations of degenerate barionic
matter, we assume an absence of transversal stresses and the
transference of masses in $V_{4}$
\begin{equation}
\label{eq:2.30}
\begin{array}{l}
T^{1}_{1}=T^{2}_{2}=T^{3}_{3}=-\widetilde{P}(\widetilde{r}), \quad
T^{0}_{0}=-\widetilde{\rho}(\widetilde{r}),
\end{array}
\end{equation}
where $\widetilde{P}(\widetilde{r})$ and
$\widetilde{\rho}(\widetilde{r})$ $\quad (\widetilde{r} \in
\widetilde{R}^{3})$ are taken to denote the internal pressure and
macroscopic density of energy defined in proper frame of reference
that is being used.
 The equations of gravitation $(a_{0})$ and ID $(\bar{a})$ fields can be obtained in Feynman
gauge~\citep{b50} as
\begin{equation}
\label{eq:2.31}
\begin{array}{l}
\Delta a_{0}=\frac{1}{2} \left\{ \widetilde{g}_{00}\frac{\partial
\widetilde{g}{}^{00}}{\partial a_{0}}
\widetilde{\rho}(\widetilde{r})- \right.\\\\\left.\left[
\widetilde{g}_{33}\frac{\partial \widetilde{g}{}^{33}}{\partial
a_{0}}+ \widetilde{g}_{11}\frac{\partial
\widetilde{g}{}^{11}}{\partial a_{0}}+
 \widetilde{g}_{22}\frac{\partial
\widetilde{g}{}^{22}}{\partial a_{0}} \right]
\widetilde{P}(\widetilde{r})\right\}, \\\\
\left(\Delta -\lambda^{-2}_{a}\right) \bar{a} =\frac{1}{2} \left\{
\widetilde{g}_{00}\frac{\partial \widetilde{g}{}^{00}}{\partial
\bar{a}} \widetilde{\rho}(\widetilde{r}) -\right.\\\\\left.\left[
\widetilde{g}_{33}\frac{\partial \widetilde{g}{}^{33}}{\partial
\bar{a}} +
 \widetilde{g}_{11}\frac{\partial
\widetilde{g}{}^{11}}{\partial \bar{a}}+
\widetilde{g}_{22}\frac{\partial \widetilde{g}{}^{22}}{\partial
\bar{a}} \right] \widetilde{P}(\widetilde{r}) \right\},
\end{array}
\end{equation}
where a diffeomorphism $\widetilde{r}(r):M_{4}\rightarrow V_{4}$ is
given $r=\widetilde{r}-R_{g}/4$. A distortion of the basis
$\widetilde{e}$ in the ID regime, in turn, yields the
transformations of Poincar\'e generators of translations. Given an
explicit form  of distorted basis vectors, it is straightforward to
derive the laws of phase transition for individual particle found in
the ID-region ($x_{0}\equiv {\ae} a_{0}=0,\,\, \bar{x} \equiv {\ae}
\bar{a}\neq 0$) of the space-time continuum, where a coupling
constant ${\ae}$ relates to Newton gravitational constant $G$ as
${\ae}^{2}=8\pi G/c^{4}$. The Poincar\'e generators $P_{\mu}$ of
translations undergone following transformations~\citep{b50,gago3}:
\begin{equation}
\label{eq:2.32}
\begin{array}{l}
\widetilde{E}=E, \quad \widetilde{P}_{1,2}=P_{1,2}\cos
\widetilde{\theta}_{3},\quad \widetilde{P}_{3}=P_{3} -\\\tan
\widetilde{\theta}_{3} \,mc, \widetilde{m}= \left| \left( m-
\tan\widetilde{\theta}_{3}\,\frac{P_{3}}{c}\right) ^{2}
+\right.\\\left.
\sin^{2}\widetilde{\theta}_{3}\,\frac{P_{1}^{2}+P_{2}^{2}}{c^{2}}
-\tan^{2} \widetilde{\theta}_{3} \,\frac{E^{2}}{c^{4}}
\right|^{\frac{1}{2}} ,
\end{array}
\end{equation}
where $E,{\vec{P}}, m$ and $\widetilde{E}, \widetilde{{\vec{P}}},
\widetilde{m}$ are ordinary and distorted energy, momentum and mass
at rest, $\tan \widetilde{\theta}_{3}=-\bar{x}$. Hence the matter
found in the ID-region $(\bar{a}\neq 0)$ of space-time continuum has
undergone phase transition of II-kind, i.e., each particle goes off
from the mass shell\,-\, a shift of mass and energy-momentum spectra
occurs upwards along the energy scale. The matter in this state is
called {\it proto-matter} with the thermodynamics differed
drastically from the thermodynamics of ordinary compressed matter.
The resulting deformed metric on $V_{4}$ in holonomic coordinate
basis takes the form
\begin{equation}
\label{eq:2.33}
\begin{array}{l}
\widetilde{g}_{00}=(1-x_{0})^{2}+\bar{x}{}^{2}, \quad
\widetilde{g}_{\mu\nu}=0 \quad (\mu \ne \nu), \\\\
\widetilde{g}_{33}=-\left[(1+x_{0})^{2}+\bar{x}{}^{2}\right],\quad
\widetilde{g}_{11}=-\widetilde{r}^{2},\\\\
\widetilde{g}_{22}=-\widetilde{r}^{2}\sin^{2}\widetilde{\theta}.
\end{array}
\end{equation}
The metric~(\ref{eq:2.33}) clearly shows  that:\, {\it a singularity
at intersection of proto-matter disk  with the event horizon (at
$x_{0}=1$) disappears  where a massive component of ID-field is not
zero ($x\neq 0$), and hence the crossing event horizon from inside
of black hole at such conditions is allowed}.

\subsection{Rational}
\begin{figure}[t]
\vspace{-1mm} \hspace{-0.4truecm}
\includegraphics[width=8cm]{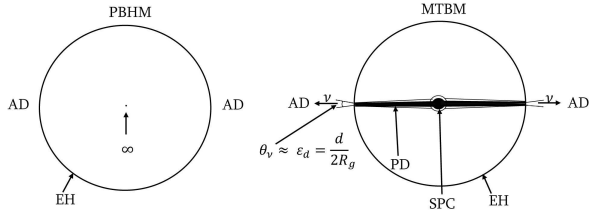}
\caption[...]{{\it Left panel}:  Phenomenological model of
non-spinning black hole. The meaningless singularity occurs at the
center inside the black hole. {\it Right panel}: Microscopic model
of non-spinning black hole, with the central stable SPC. An
infalling matter with the time forms PD around the SPC. In final
stage of growth, a PD has reached out the edge of the event horizon.
Whereas a metric singularity inevitably disappears and UHE neutrinos
may escape from event horizon to outside world through vista, i.e. a
thin belt area $S=2\pi R_{g}d$ of intersection of proto-matter disk
with the event horizon,  with opening angle $\theta_{\nu}$.

Accepted notations: EH=Event Horizon, AD=Accretion Disk,
SPC=Superdense Proto-matter Core, PD=Proto-matter Disk.} \label{F1}
\end{figure}

Consisting of the proto-matter core and the outer layers of ordinary
matter, the SPC configuration is the spherical-symmetric
distribution of barionic matter in many-phase stratified states. A
layering is a consequence of the onset of different regimes in
equation of state. Each configuration is defined by the two free
parameters of central values of particle concentration
$\widetilde{n}(0)$ and dimensionless potential of space-like
ID-field $\bar{x}(0)$. The interior gravitational potential
$x^{int}_{0}(r)$ matches into the exterior one $x^{ext}_{0}(r)$ at
the surface of the configuration. The central value of the
gravitational potential $x_{0}(0)$ can be found by reiterating
integrations when the sewing condition of the interior and exterior
potentials holds. The simulations confirm in brief the following
scenario. The energy density and internal pressure have sharply
increased in proto-matter core, with respect to corresponding
central values of neutron star, proportional to gravitational forces
of compression. This counteracts the collapse and equilibrium holds
even for the masses up to $\sim 10^{10}M_{\odot}$. Encapsulated in a
complete set of equations of SPC-configuration, the SPC is a robust
structure that has stood the tests of the most rigorous theoretical
scrutinies of its stability, which  was a central issue
in~\citep{b501}. Minimizing the total energy gives the equilibrium
configurations. The second derivative of total energy gives
stability information. Although a relativity tends to destabilize
configurations, however, a numerical integrations of the stability
equations of SPC clearly proves the stability of resulting cores.
Due to it, the stable equilibrium holds as well in outward layers
and, thus, an accumulation of matter is now allowed around the
stable SPC. To emphasize the distinction between phenomenological
and microscopic black hole models, we present their schematic plots
in Fig.~\ref{F1}, to guide the eye.
%----------------------------------------------------------------
%-------------------------------------------------------------
%----------------------------------------------------------------
One of the most remarkable drawback of MTBH is that the central
singularity cannot occur, which is replaced by finite though
unbelievably extreme conditions held in the stable SPC where,
nevertheless, static observers are existed. The SPC is always found
inside the event horizon sphere, therefore it could be observed only
in presence of accreting matter. The SPC surrounded by the accretion
disk presents the microscopic model of AGN. The SPC accommodates the
highest energy scale up to hundreds ZeV in central proto-matter core
of suppermasive black holes, which accounts for the spectral
distribution of the resulting radiation of galactic nuclei. An
external physics of accretion onto the black hole in the earlier
half of its lifetime is identical to the processes in
Schwarzschild's model. A crucial difference in the model context
between the phenomenological and microscopic black holes comes in
when one looks for the spontaneous breaking of gravitation gauge
symmetry at huge energies, and thereof making room for growth and
merging behavior of black holes. We have shown that the AGN
evolution and black hole growth, when properly analysed, agree with
our claims.

\subsection{The intermediate mass,  pre-radiation time and  initial redshift of seed
black hole}  We argue that in the framework of MTBH, the black hole
seeds might grow driven by the accretion of outside matter when they
were getting most of their masses. An infalling matter with time
forms a thin proto-matter disk around the proto-matter core tapering
off faster at reaching out the event horizon. As a  a metric
singularity inevitably disappears and the ZeV-neutrinos, produced
via simple or modified URCA\footnote{G. Gamow was inspired to name
the process URCA after the name of a casino in Rio de Janeiro, when
M. Schenberg remarked to him that "the energy disappears in the
nucleus of the supernova as quickly as the money disappeared at that
roulette table"} processes in deep layers of SPC and proto-matter
disk, may escape from event horizon to outside world through a thin
belt area $S=2\pi R_{g}d$, even after the strong neutrino trapping.
The thickness of proto-matter disk at the edge of event horizon is
denoted by $d$. The neutrinos are collimated in very small opening
angle $\theta_{\nu}\simeq \varepsilon_{d}=\frac{d}{2R_{g}}\ll 1$.
The {\em trapping} is due to the fact that as the neutrinos are
formed in proto-matter medium at super-high densities they
experience greater difficulty escaping from it before being dragged
along with the matter, i.e. the neutrinos are {\em trapped} comove
with matter. In this framework we introduced a notion of
pre-radiation time (PRT) of black hole which is referred to as a
lapse of time $T_{BH}$ from the birth of black hole till neutrino
radiation \,-\,provision for the earlier half of the lifetime of
black hole:
\begin{equation}
\begin{array}{l}
T_{BH}=\FFr{M_{d}}{\dot{M}}.
\end{array}
\label{R0}
\end{equation}
Here $M_{d}$ is the mass of proto-matter disk, $\dot{M}$ is an
appropriately averaged mass accretion rate. To render our discussion
here a bit more transparent, we present the relation between typical
PRT versus bolometric luminosity of suppermassive black holes as
follows:
\begin{equation}
\begin{array}{l}
T_{BH} \simeq
0.32\FFr{R_{d}}{r_{OV}}\left(\FFr{M_{BH}}{M_{\odot}}\right)^{2}\FFr{10^{39}W}{
L_{bol}}\,\mathrm{yr},
\end{array}
\label{R1}
\end{equation}
where $R_{d}$ is the radius of the proto-matter core,
$r_{OV}=13.68\,{\rm km}$. At times $>T_{BH}$, the black hole no
longer holds as a region of spacetime that cannot communicate with
the external Universe. In this framework, we computed the fluxes of
ZeV-neutrinos from plausible accreting supermassive black holes
closely linking with the 377 AGNs~\citep{gago2,gago3}. In accord,
the AGNs are favored as promising pure neutrino sources because the
computed neutrino fluxes are highly beamed along the plane of
accretion disk, and peaked at high energies and collimated in very
small opening angle $\theta_{\nu}\sim \varepsilon_{d}\ll 1$. While
hard to detect, the extragalactic ZeV-neutrinos may reveal clues on
the puzzle of the origin of ultra-high energy cosmic rays, as they
have the advantage of representing unique fingerprints of hadron
interactions and, therefore, can initiate the cascades of
UHE-particles with energies exceeding $1.0\times10^{20}$ eV
(comprehensive reviews can be found
in~\citet{Cast,Let,Sigl,KO,Sem}).

Consequently, we have studied a growth of proto-matter disk and
derived the mass of black hole seed ~\citep{gago3}\,-\,provision for
the second half of the lifetime of black hole:
\begin{equation}
\begin{array}{l}
\FFr{M_{BH}^{Seed}}{M_{\odot}}\simeq
\FFr{M_{BH}}{M_{\odot}}\left(1-2.305\,\FFr{R_{d}}{r_{OV}}\,\FFr{M_{BH}}{M_{\odot}}\right),
\end{array}
\label{R2}
\end{equation}
and initial redshift
\begin{equation}
\begin{array}{l}
z^{Seed}\simeq z+ H_{0}\,T_{BH},
\end{array}
\label{R2Z}
\end{equation}
where $H_{0}$ is the Hubble's constant. Whereas interpreting the
redshift as a cosmological Doppler effect, and that the Hubble law
could most easily be understood in terms of expansion of the
Universe, as a further consistency check, we then turned the problem
around by asking the purely academic question of principle what
could be the initial redshift, $z^{Seed}$, of seed black hole if the
mass, the luminosity and the redshift, $z$, of black hole at present
time are known. Equation~(\ref{R2}) holds at
$\frac{R_{d}}{r_{OV}}\leq 0.023\,\frac{R^{s}_{g}}{r_{OV}}$, where
$R^{s}_{g}$ is the gravitational radius of black hole seed. We have
undertaken a large series of numerical simulations with the goal to
trace an evolution of the mass assembly history of 377 accreting
supermassive black hole seeds in AGNs to the present time and
examine the observable signatures today. Given  the redshifts,
masses and luminosities of these black holes at  present time
collected from the literature, we compute the initial redshifts and
masses of the corresponding black hole seeds.

Having gained some insight into the supermassive black hole physics,
let us now comment briefly on the implications for IMBHs.
Equation~(\ref{R2}) shows that the seed of IMBH can be formed in the
stellar mass black hole region if the condition
\begin{equation}
\begin{array}{l}
\FFr{R_{d}}{r_{OV}}\leq
0.434\,\FFr{M_{\odot}}{M_{BH}}\left(1-\FFr{M_{\odot}}{M_{BH}}\right)
\end{array}
\label{R222}
\end{equation}
holds. A creation of IMBHs may be possible in the young stellar
populations that we generally find ULXs co-habiting with. The IMBH
with mass, say $M_{BH}\simeq 1.0\times 10^{3}\,M_{\odot}$, may
originate from the seed with stellar mass of $\sim 1\,M_{\odot}$ at
$R_{d}/r_{OV}\simeq 4.3\times 10^{-4}$, and will reach to the finale
state after a lapse of growing time of order $T_{BH}\simeq
M_{BH}/\dot{M}$.

\section{The mass-exchange binary containing
M82X-2} As in the case of neutron stars, we expect that accreting
black holes are fast spinning objects. For the self-contained
arguments, we need to extend the preceding algorithm of non-spinning
MBHM to its spinning counterpart, which is almost a matter of
routine, to change the geometry of static SPC to a more general one,
describing axisymmetric rotating SPC. For the standard calculations
of rapidly rotating relativistic bodies in astrophysics, reader may
refer to~(e.g.~\citet{Cart,Bard,KEH,CST1,CST2,BGS,Bon,Ster}) and
references therein.

\subsection{The geometry of rotating
axisymmetric SPC} In this subsection we will collect together the
results which are required later. The non-spinning SPC is static and
spherically symmetric. So, one needs to be clear about more general
geometry which can describe rotating axisymmetric SPCs. The
principle foundation of the spinning configurations first comprises
the following additional distinctive features with respect to
non-spinning ones: 1) Rapid rotation causes the shape of the SPC to
be flattened by centrifugal forces\,-\, flattened at poles and
bulged at equator (oblate spheroid, which is second order effect in
the rotation rate).  2) A rotating massive SPC drags space and time
around with it.  The local inertial frames are dragged by the
rotation of the gravitational field, i.e. a gyroscope orbiting near
the SPC will be dragged along with the rapidly rotating SPC. This is
probably the most remarkable feature that could serve as a link with
the general description of spacetime (also see~\citet{gi}). Beside
the geodetic procession, a spin of the body produces in addition the
Lense-Thirring procession. To look into the future, measurement of
the gyrogravitational ratio of particle would be a further step, see
e.g. \citep{Ni10} and references therein, towards probing the
microscopic origin of gravity. Let the world coordinate $t(= x^{0})$
be the time (in units of c), and $\phi(=x^{1})$ be the azimuthal
angle about the axis of symmetry. Moreover, a metric of a two-space
with positive or negative signature, $(x^{2},x^{3})$, can always be
brought to the diagonal form by a more coordinate transformation.
Since the source of gravitational field has motions that are pure
rotational about the axis of symmetry, then the energy-momentum
tensor as the source of the metric will have the same symmetry.
Namely, the space ${\mathcal{M}}_{4}$ would be invariant against
simultaneous inversion of time $t$ and azimuthal angle $\phi$. The
$3+1$ formalism is the most commonly used approach in which, as
usual, spacetime is decomposed into the one parameter family of
space-like slices\,-\, the hypersurfaces $\Sigma_{t}$. The study of
a dragging effect is assisted by incorporating with the soldering
tools in order to relate local Lorentz symmetry to curved spacetime.
These are the linear frames and forms in tangent fiber-bundles to
the external general smooth differential manifold, whose components
are so-called tetrad (vierbein) fields. Whereas, the
${\mathcal{M}}_{4}$ has at each point a tangent space,
${T}_{{x}}{\mathcal M}_{4}$, spanned by the anholonomic orthonormal
frame field, ${e}$, as a shorthand for the collection of the
4-tuplet $({e}_{0}=\exp (-\nu)\,(\partial_{t}+\omega
\partial_{\phi}),\,\, {e}_{1}=\exp (-\psi)\,\partial_{\phi},
\,\,{e}_{2}=\exp (-\mu_{2})\,\partial_{2}, \,\,{e}_{3}=\exp
(-\mu_{3})\,\partial_{3})$, where
${e}_{a}={e}_{a}^{\phantom{a}\mu}\,{\partial}_{\mu}$,
$\,\,{e}_{a}^{\phantom{a}\mu}$ is the the soldering form between the
tangent space and the spacetime manifold. This is called a Bardeen
observer, locally nonrotating observer, or the local Zero
Angular-Momentum Observers (ZAMO), i.e. observers whose worldlines
are normal to the hypersurfaces defined by constant coordinate time,
$t=const$, also called Eulerian observers. Here we use Greek
alphabet ($\mu, \nu, \rho,...= 0, 1, 2, 3$) to denote the holonomic
world indices related to ${\mathcal{M}}_{4}$, and the first half of
Latin alphabet ($a, b, c, . . . = 0, 1, 2, 3$) to denote the
anholonomic indices related to the tangent space. The frame field,
${e}$, then  defines a dual vector, ${\vartheta}$, of differential
forms, $ {\vartheta}=\left(
                    \begin{array}{c}
                      {\vartheta}{}^{0} =\exp \nu\,dt\\
                      {\vartheta}{}^{1}=\exp \psi\,(d\psi-\omega dt)\\
                      {\vartheta}{}^{2}=\exp \mu_{2}\, dx^{2}\\
                      {\vartheta}{}^{3}=\exp \mu_{3}\, dx^{3}\\
                    \end{array}
                  \right),
$ as a shorthand for the collection of the
${\vartheta}{}^{b}={e}{}^{b}_{\phantom{a} \mu}\,d{x}{}^{\mu}$, whose
values at every point form the dual basis, such that
${e}_{a}\,\rfloor\, {\vartheta}{}^{b}=\delta^{b}_{a}$, where
$\rfloor$ denoting the interior product, namely, this is a
$C^{\infty}$-bilinear map $\rfloor:\Omega^{1}\rightarrow \Omega^{0}$
with $\Omega^{p}$ denotes the $C^{\infty}$-modulo of differential
p-forms on ${\mathcal{M}}_{4}$. In components
${e}_{a}^{\phantom{a}\mu}\,{e}{}^{b}_{\phantom{a}\mu}=\delta^{b}_{a}$.
The norm $ds$ of infinitesimal displacement $d{x}^{\mu}$ on
${\mathcal{M}}_{4}$, describing the stationary and axisymmetric
spacetimes, reads
\begin{equation}
\begin{array}{l}
ds:\phantom{a}={e}\,{\vartheta}={e}_{\mu}\otimes{\vartheta}{}^{\mu}
\,\in\,{\mathcal{M}}_{4}.
\end{array}
\label{S0}
\end{equation}
Therefore, the holonomic metric on the space ${\mathcal{M}}_{4}$ can
be recast into the form
\begin{equation}
\begin{array}{l}
{g}={g}_{\mu\nu}\,{\vartheta}{}^{\mu}\otimes{\vartheta}{}^{\nu}=
{g}({e}_{\mu}, \,{e}_{\nu})\,
{\vartheta}{}^{\mu}\otimes{\vartheta}{}^{\nu},
\end{array}
\label{M0}
\end{equation}
with the components ${g}_{\mu\nu}={g}({e}_{\mu}, {e}_{\nu})$ in dual
holonomic basis $\{{\vartheta}{}^{\mu}\equiv d{x}{}^{\mu}\}$. That
is
\begin{equation}
\begin{array}{l}
ds^{2}=-\exp (2\nu)dt^{2}+\exp (2\psi)(d\phi-\omega
dt)^{2}+\\\exp(2\mu_{2})(dx^{2})^{2}+\exp(2\mu_{3})(dx^{3})^{2},
\end{array}
\label{M1}
\end{equation}
where the five quantities $\nu,\psi,\omega,\mu_{2}$ and $\mu_{3}$
are only functions of the coordinate $x^{2}$ and $x^{3}$. In the
case at hand, the metric function $\omega$ is the angular velocity
of the local ZAMO with respect to an observer at rest at infinity.
Thereby the redshift factor $\alpha\equiv\exp \nu$ is the time
dilation factor between the proper time of the local ZAMO and
coordinate time $t$ along a radial coordinate line, i.e. the
redshift factor for the time-slicing of a spacetime. In accord, all
the geometrical objects are split into corresponding components with
respect to this time-slice of spacetime. In particular, the
splitting of manifold ${\mathcal{M}}_{4}$ into a foliation of
three-surfaces will induce a corresponding splitting of the affine
connection, curvature and, thus, of the energy-momentum tensor which
can as well be written in terms of the energy density
$E=T(n,n)=T_{\mu\nu}n^{\mu}n^{\nu}$ measured by an adapted Eulerian
observer of four-velocity $n^{\mu}$, the momentum flow
$J_{\alpha}=-\gamma^{\mu}_{\alpha}T_{\mu\nu}n^{\nu}$ and the
corresponding stress tensor
$S_{\alpha\beta}=\gamma^{\mu}_{\alpha}\gamma^{\nu}_{\beta}T_{\mu\nu}\,\,
(S=S^{\alpha}_{\alpha})$, that is
$T^{\alpha\beta}=En^{\alpha}n^{\beta}+n^{\alpha}J^{\beta}+J^{\alpha}n^{\beta}+S^{\alpha\beta}$.
Here $n^{\alpha}$ is the unit orthogonal vector to the hypersurface
$\Sigma_{t}$, whereas the spacetime metric $g$ induces a first
fundamental form with the spatial metric $\gamma_{\alpha\beta}$ on
each $\Sigma_{t}$ as $\gamma_{\alpha\beta}=g_{\alpha\beta}+
n_{\alpha}n_{\beta}$. The form (\ref{M1}) includes one gauge freedom
for the coordinate choice. For the spherical type coordinates
$x^{2}=\theta$ and $x^{3}=r$, for example, so-called quasi-isotropic
gauge corresponds to $\gamma_{r\theta}=0$ and
$\gamma_{\theta\theta}=r^{2}\gamma_{rr}$. Then, one may define the
second fundamental form which associates with each vector tangent to
$\Sigma_{t}$, and the extrinsic curvature of the hypersurface
$\Sigma_{t}$ as minus the second fundamental form.  Aftermath, one
can define the usual Lorentz factor $W=-n_{\mu}u^{\nu}=\alpha u^{t}$
for a perfect fluid with conventional stress-energy tensor
$T^{\mu\nu}=(\rho+P)u^{\mu}u^{\nu}+Pg^{\mu\nu}$, where $\rho$ is the
energy density and $P$ is the pressure. Hence $E=W^{2}(\rho+P)-P$
and $J^{i}=(E+P)v^{i}$, where the fluid three-velocity $v^{i}
(i=1,2,3)$ implies $u^{i}=W(v^{i}-\beta^{i}/\alpha)$. Thereby the
resulting stress tensor can be written
$S_{ij}=(E+P)v^{i}v_{j}+P\gamma_{ij}$. The four-velocity for
rotating fluid reads $u=u^{i}(\partial/\partial t)+\Omega
\partial/\partial \phi$, where $\Omega=u^{\phi}/u^{t}$ is the fluid
angular velocity as seen by an inertial observer at rest at
infinity. It is convenient to give the equations in the isotropic
gauge, $\exp (2\mu_{2})=\exp (2\mu_{3})=\exp (2\mu)$, in the
cylindrical coordinates $d\varrho=dx^{2}, \quad dz=dx^{3}$, or in
pseudospherical coordinates  $r\,d\theta=dx^{2}$ and $dr=dx^{3}$,
where the cylindrical radius is written $\exp (\psi)=r\,\sin
\theta\, B\,\exp (-\nu)$, with $B$ denoting a function of $r$ and
$\theta$ only. Therefore, the metric~(\ref{M1}) becomes
\begin{equation}
\begin{array}{l}
ds^{2}=-\exp (2\nu)dt^{2}+B^{2}r^{2}\sin^{2}\theta\,\exp
(-2\nu)(d\phi-\\\omega dt)^{2}+\exp(2\mu)(dr^{2}+r^{2}d\theta^{2}).
\end{array}
\label{M2}
\end{equation}
Consequently, the components of the energy - momentum tensor of
matter with total density $\rho$ and pressure $P$ are given in the
non-rotating anholonomic orthonormal frame as $T^{ab}=e^{a}_{\mu}
e^{b}_{\nu}T^{\mu\nu}, \quad T^{00}=W^{2}(\rho+PV^{2}), \quad
T^{11}=W^{2}(\rho+PV^{2}),\quad T^{01}=W^{2}(\rho+P)V$ and
$T^{22}=T^{33}=P$, with its trace $T=-\rho+3P$, where  $V$ is the
velocity (in units of c) with respect to the Bardeen observer
$V=\frac{\varrho B (\Omega-\omega)}{\alpha^{2}}$, so
$W=\frac{1}{\sqrt{1-V^{2}}}$. However, at this point we cut short
and, in what follows, we will refrain from providing further lengthy
details of the mathematical apparatus of proposed gravitation theory
at huge energies and rigorous solution of the extended equations
describing the spinning MBHM, which is beyond the scope of this
report. This entire set includes the gravitational and ID field
equations of the time-slices generated by the above stress-energy
tensor, the angular momentum equation with the momentum flow
determining the frame-dragging potential, the hydrostatic
equilibrium equation and the state equation specified for each
domain of many layered spinning SPC- configurations. Theoretical
evolutionary paths of this type which are suitable for rigorous
comparison with the behavior of observed sources will be separate
topic of comprehensive investigation elsewhere. But some evidence
for a simplified physical picture, without loss of generality, is
highlighted in the rest of this section where we extend preceding
developments of MTBH in concise form, without going into the
subtleties, as applied to the initially rigid-body spinning IMBH
configuration of angular velocity $\Omega$.
%----------------------------------------------------------------
%-------------------------------------------------------------
\begin{figure*}
\vspace{-10mm} \hspace{1.5truecm}
\includegraphics[width=14cm]{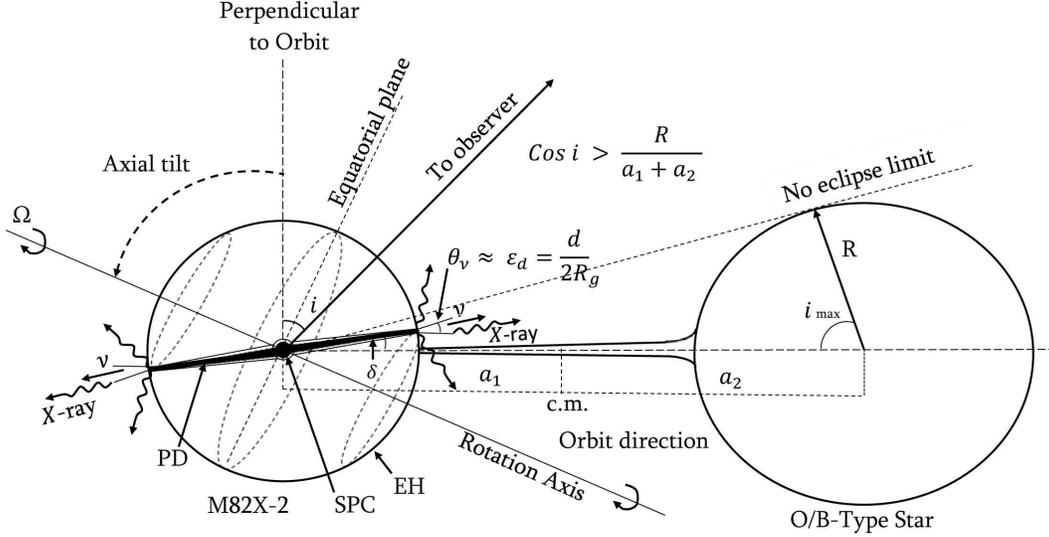}
\caption[...]{A schematic SIMBH model of M82X-2 constituting
mass-exchange binary with the O/B-type donor star. The angle $i$ is
the binary inclination with respect to the plane of the sky. No
eclipse condition holds. In final stage of growth, PD has reached
out the edge of the EH. The thermal defuse blackbody X-rays beams
may escape from SIMBH through a thin belt area $S=2\pi R_{g}d$ to
outside world that sweep past Earth like lighthouse beacons.
Parameters of a binary system is viewed in the orbital plane. The
picture is not to scale. Accepted notations: EH=Event Horizon,
SPC=Superdense Proto-matter Core, PD=Proto-matter Disk.} \label{F2}
\end{figure*}
%----------------------------------------------------------------

\subsection{The X-ray radiation from SIMBH} If this
were the case, eventually the spinning proto-matter core and a thin
co-spinning proto-matter disk driven by accretion would be formed.
The evolution and structure of a proto-matter disk is largely
determined by internal friction. Before tempting to build a physical
model in quest, the other features of SIMBH configuration also need
to be accounted which comprise the whole of the case. The fact that
the rotational energy has a steeper dependence on the radius of the
compact object than the internal energy in the relativistic limit is
quite significant. Equilibrium can always be achieved for massive
configurations with nonzero angular momentum by decreasing its
radius. Also, there are two characteristic features that distinguish
a spinning relativistic SPC-configuration from its non-spinning
counterpart: 3) The geodetic effect, as in case of a gyroscope,
leads an accretion stream to a tilting of its spin axis in the plain
of the orbit. Hence  a proto-matter disk will be tilted from the
plane of accretion on a definite angle $\delta$ towards the equator.
4) Besides the UHE neutrinos, produced in the deep interior layers
of superdense proto-matter medium as in case of non-spinning model,
the additional thermal defuse blackbody radiation is released from
the outer surface layers of ordinary matter of spinning SPC and
co-spinning proto-matter thin disk. All of the rotational kinetic
energy is dissipated as thermal blackbody radiation. This is due to
the physical condition that these layers optically thick and,
eventually, in earlier half of the lifetime of spinning black hole,
at times $< T_{BH}$, the strict thermodynamic equilibrium prevails
for this radiation because there would be no net flux to outside of
event horizon in any direction. That is, the emission from the
isothermal, optically thick outer layers at surface is blackbody,
which is the most efficient radiation mechanism. This radiation is
free of trapping. With this guidelines to follow, we may proceed in
relatively simple way toward first look at some of the associated
physics  and can be quick to estimate the physical characteristics
of mass-exchange X-ray binaries. Examining the pulsations revealed
from M82X-2, as a working model we assume the source  of the flashes
to be a self-gravitating SIMBH resided in the final stage of growth.
This implies that a thin co-spinning proto-matter disk  has reached
out the edge of the event horizon. Due to it a metric singularity
inevitably disappears at the intersection of proto-matter disk with
the event horizon.  Hence, the crossing event horizon at such
conditions from inside of black hole  is allowed. This general
behavior is very robust and depends very little on the details of
the model of SPC. A principle physical properties of this phenomena
for non-rotating SPC are already discussed in subsection 2.3.
Without being carefully treated, even though these properties are of
great significant for a rotating SPCs too. Actually, by virtue of
subsect.3.1, they will retain for a more general geometry if
metric~(\ref{eq:2.33}) will appropriately be recast into~(\ref{M2})
type, which describes a rotating axisymmetric SPCs. We conclude on
the basis of these observations that the energy can be carried away
to outside world by the thermal defuse blackbody X-ray radiation
through a thin belt area $S$. As M82X-2 spins, the pulse profile
results from the axial tilt or obliquity. Hence the X-ray beams
intercept Earth-like lighthouse beacons. The orbital motion causes a
modulation in the observed pulse frequency. The SIMBH model of
M82X-2 in binary system is schematically plotted in Fig.~\ref{F2}.
No eclipse condition holds. The parameters of a binary system is
viewed in the orbital plane.

\subsection{Basic geometry:\, Implications on the pulse profile and mass scaling}
\begin{figure*}
\vspace{-20mm} \hspace{3truecm}
\includegraphics[width=9cm]{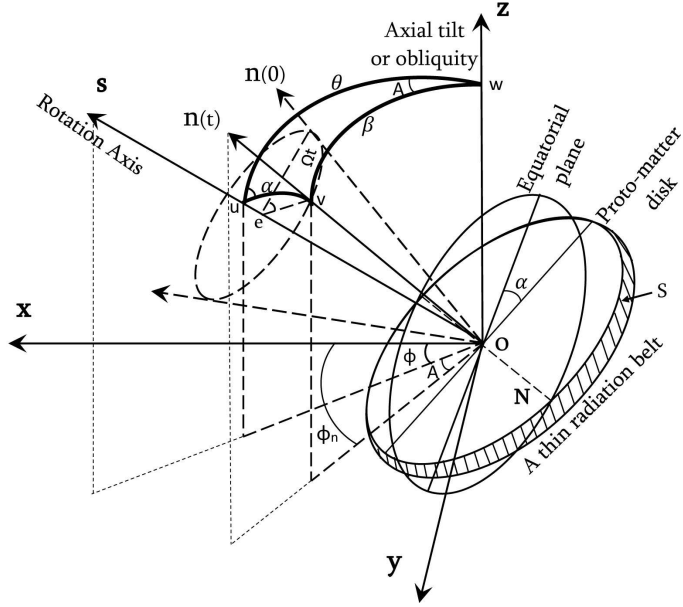}
\caption[...]{A schematic plot explaining the spherical triangle
solved by the law of cosines.  The space-fixed Cartesian coordinate
system is labeled (z,x,y), with zx as a plane-of sight. Axis $s$ of
the M82X-2-fixed frame is rotation axis. The angles $\theta$ and
$\phi$ are spherical polar coordinates. The line of nodes N is
defined as the intersection of the equatorial and proto-matter disk
planes. It is perpendicular to both the z axis and vector
$\vec{n}(t)$, where $\vec{n}(t)$ is the normal to the plane of
proto-matter disk at moment $t$. The $\vec{n}(0)$ lies in the plane
of zs.  The lengths of three sides of a {\em spherical triangle}
(shown at top) are $\theta=\widehat{(z,s)} $,
$\alpha=\widehat{(s,n)}$ and $\beta=\widehat{(z,n)}$.  The vertex
angle of opposite $\beta$ is $\Omega t$.} \label{F3}
\end{figure*}

\begin{figure*}
\vspace{-20mm} \hspace{3truecm}
\includegraphics[width=11cm]{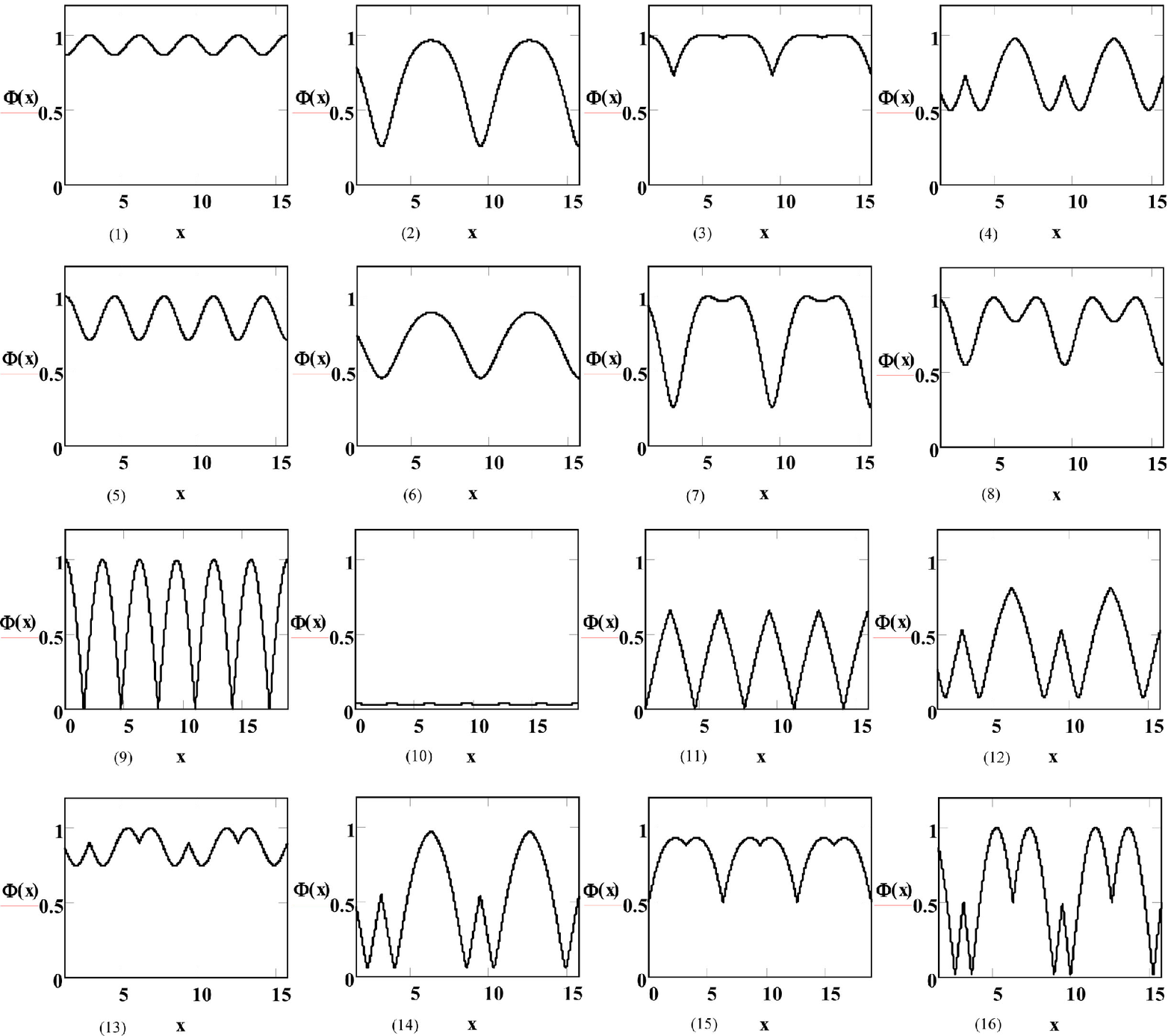}
\caption[...]{Characteristic phase profiles versus the time
($x\equiv \Omega t$),  viewed at the following position angles
$(\theta, \alpha,\phi)$: \,\, (1)\,\,$(45^{0}, 30^{0}, 0^{0})$;
(2)\,\,$(45^{0}, 30^{0}, 90^{0})$; (3)\,\,$(45^{0}, 30^{0},
135^{0})$; (4)\,\,$(45^{0}, 30^{0}, 60^{0})$; (5)\,\,$(45^{0},
90^{0}, 90^{0})$; (6)\,\,$(45^{0}, 18^{0}, 90^{0})$;
(7)\,\,$(45^{0}, 60^{0}, 90^{0})$; (8)\,\,$(45^{0}, 78.3^{0},
90^{0})$; (9)\,\,$(90^{0},90^{0}, 180^{0})$; (10)\,\,$(90^{0},
0.6^{0}, 72^{0})$; (11)\,\,$(90^{0}, 30^{0}, 60^{0})$;
(12)\,\,$(72^{0}, 30^{0}, 60^{0})$; (13)\,\,$(60^{0}, 90^{0},
60^{0})$; (14)\,\,$(60^{0}, 45^{0}, 60^{0})$; (15)\,\,$(60^{0},
153^{0}, 60^{0})$; (16)\,\,$(1.2^{0}, 90^{0}, 60^{0})$. } \label{F4}
\end{figure*}

To see where all this is leading to, let us consider next the real
issue that of the physical description of M82X-2. A knowledge of the
dynamical mass measurements of the compact objects that power ULXs
is a primarily necessary prerequisite to the derivation of a
complete picture about the physical nature of ULXs. Keeping in mind
aforesaid, we are now in a position to derive a general pulse
profile dependent upon the position angles, and give a quantitative
account of a potential dynamical mass scaling of M82X-2 and other
energetics. The most reliable method is to measure the mass function
through the secondary mass and orbital parameters, which can be
measured only if the secondary donor star is optically identified.
In the absence of direct mass-function measurements from
phase-resolved optical spectroscopy, we still have to rely on X-ray
spectral and timing modeling and other indirect clues. To bring this
point sharply into focus note that in case of the first
ultraluminous pulsar, only the X-ray mass function is
measured~\citep{NuS}, when the optical secondary is unknown and most
of the orbital parameters are yet to be measured. However, exploring
the key physical characteristics of a SIMBH model, let us consider
the space-fixed Cartesian coordinate system labeled (z,x,y), with zx
as a plane-of sight, and the axis $s$ of the M82X-2-fixed frame as
the rotation axis. A schematic plot is given in Fig.~\ref{F3}. Here
and throughout we now use following notational conventions. The
angles $\theta$ and $\phi$ are spherical polar coordinates. The
observed pulses are produced because of periodic variations with
time of the projection on the plane-of sight, $d_{zx}(t)$, of the
vector $\vec{d}(t)$ collinear to $\vec{n}(t)$ ($\vec{d}(t)=d\,
\frac{\vec{n}(t)}{|n(t)|}$), where $\vec{n}(t)$ is the normal to the
plane of the proto-matter disk at the moment $t$. The $\vec{n}(0)$
lies in the plane of zs. So, these pulsations are due to the fact
that the visible surface is less at one moment than at another.
%----------------------------------------------------------------
%-------------------------------------------------------------
%----------------------------------------------------------------
The Fig.~\ref{F3} explains the {\em spherical triangle}. That is,
given a unit sphere, a {\em spherical triangle} on the surface of
the sphere is defined by the great circles connecting three points
u, v, and w on the sphere (shown at top). The lengths of these three
sides are $\alpha=\widehat{(s,n)}$ (from u to v), the axial tilt
$\theta=\widehat{(z,s)}$ (from u to w), and $\beta=\widehat{(z,n)} $
(from v to w). The angles of the corners $u$ and $e$ opposite
$\beta$ equal $u=e=\Omega t$. The proto-matter disk was shifted from
the orbital direction on angle $\delta=\theta-\alpha $ towards the
equator. The projection $d_{zx}(t)$ is then written
\begin{equation}
\begin{array}{l}
d_{zx}(d,\theta,\phi,\alpha,t)=\sqrt{d^{2}_{z}(d,\theta,\alpha,t)+d^{2}_{x}(d,\theta,\alpha,\phi,t)},
\end{array}
\label{R03}
\end{equation}
where
\begin{equation}
\begin{array}{l}
d_{z}(d,\theta,\alpha,t)\equiv \vec{d}(...,t)\cdot \vec{e_{z}}=d
\cos \beta (\theta,\alpha,t),
\end{array}
\label{R333}
\end{equation}
and
\begin{equation}
\begin{array}{l}
d_{x}(d,\theta,\alpha,\phi,t)\equiv \vec{d}(...,t)\cdot
\vec{e_{x}}=\\d\, \sin \beta(\theta,\alpha,t) \cos \phi_{n}(\theta,
\alpha, \phi,t).
\end{array}
\label{R333333}
\end{equation}
Here the $\vec{e}_{z}$ and $\vec{e}_{x}$ denote unit vectors along
the axes $z$ and $x$, respectively, $\phi_{n}=\phi +A$ is the
azimuthal angle of vector $\vec{n}(t)$. The vertex angle opposite
the side $\alpha$ is $A$. To the extent that all of the rotational
energy of M82X-2 is dissipated as thermal defuse X-ray blackbody
radiation, this  may escape from the event horizon to outside world
only through a thin belt area $S$. The radiation arisen from per
area of surface is $\sigma T^{4}_{s }$, where $T_{s }$ is the
surface temperature, $\sigma$ is the Stefan-Boltzmann constant.
Therefore, the pulsed luminosity $\widetilde{L}$ will be observed if
and only if the projection of the belt area $S_{zx}=2\pi
R_{g}d_{zx}(d,\theta,\phi,\alpha,t)$ on the plane-of sight zx is not
zero. So, pulsed luminosity reads
\begin{equation}
\begin{array}{l}
\widetilde{L}(R_{g},d,T_{s },\theta,\phi,\alpha,t)=S_{zx}\,\sigma
T^{4}_{s }= \\2\pi
R_{g} d_{zx}(d,\theta,\phi,\alpha,t)\,\sigma T^{4}_{s }\equiv\\
L_{0}(M,d,T_{s })\,\Phi(\theta,\phi,\alpha,t),
\end{array}
\label{R3}
\end{equation}
where the amplitude and phase, respectively, are
\begin{equation}
\begin{array}{l}
L_{0}(M,d,T_{s })\simeq 1.05\times 10^{4}\,({\rm erg
\,s^{-1}})\FFr{M}{M_{\odot}}\,\FFr{d}{\rm
m}\,\FFr{T^{4}_{s}}{K^{4}},
\\\\
\Phi(\theta,\phi,\alpha,t)\equiv
\sqrt{1-\sin^{2}\beta\sin^{2}(\phi+A)}.
\end{array}
\label{RL}
\end{equation}
The {\em spherical triangle} is solved by the law of cosines
\begin{equation}
\begin{array}{l}
\cos \beta(\theta, \alpha, t)=\cos \theta \cos \alpha +\sin \theta
\sin \alpha \cos \Omega t,
\\
\cos A(\theta, \alpha, t)=\FFr{\cos\alpha-\cos \theta \cos
\beta}{\sin \theta \sin \beta}.
\end{array}
\label{RA}
\end{equation}
Consequently, the pulsed flux can be written in the form
\begin{equation}
\begin{array}{l}
\widetilde{F}(R_{g},d,\theta,\phi,\alpha,t)=
F_{0}(M,d)\,\Phi(\theta,\phi,\alpha,t).
\end{array}
\label{RF}
\end{equation}
where, given the distance $D\simeq 3.6\,{\rm Mpc}$ to the galaxy
M82~\citep{NuS}, the flux amplitude is
\begin{equation}
\begin{array}{l}
F_{0}(M,d)=\FFr{L_{0}(M,d)}{4\pi D^{2}}\simeq\\\\
6.8\times 10^{-48}\,({\rm erg \,s^{-1}
cm^{-2}})\,\FFr{M}{M_{\odot}}\,\FFr{d}{\rm
m}\,\FFr{T^{4}_{s}}{K^{4}}.
\end{array}
\label{Rf0}
\end{equation}
Thus, the theoretical model of periodic source M82X-2 left six free
parameters: $(M,d,T_{s },\theta,\,\phi,\,\alpha)$. The figure
Fig.~\ref{F4} reveals the diversity of the behavior of
characteristic phase $\Phi(\theta,\phi,\alpha,x\equiv \Omega t)$
profiles versus the time, viewed at given position angles
$(\theta,\,\phi,\,\alpha)$.
%-------------------------------------------------------------
%-------------------------------------------------------------
The observed X-ray pulse is further determined by the complicated
transfer of X-ray photons from the surface of M82X-2 through regions
of external accreting plasma. At hard look, the position angles
being the parameters of a model function can be evaluated via
nonlinear regression analysis to the approximate solution of
overdetermined systems to best fit a data set from observed pulsed
profile of M82X-2. These missing ingredients are the shortcoming of
present framework, which will be solved by iterative refinement
elsewhere. Now according to~(\ref{R3}), for maximum value of pulsed
luminosity either at $\beta=\pi s$ or $\phi+A=\pi
s$\,\,(s=0,1,2,...), we have $L_{0}(M,d,T_{s })=\widetilde{L}(3-
30\,{\rm keV}) = 4.9 \times 10^{39} \,{\rm erg \,s^{-1}}$ or
$F_{0}(M,d)=\widetilde{F}(3- 30\,{\rm keV}) \simeq 3.16 \times
10^{-12} \,{\rm erg \,s^{-1}cm^{-2}}$. Hence the surface temperature
scales $\propto T^{-4}_{s}$ with the black hole mass:
\begin{equation}
\begin{array}{l}
\FFr{M}{M_{\odot}}=\FFr{4.66\times
10^{35}\,K^{4}}{T^{4}_{s}}\,\FFr{\rm m}{d}.
\end{array}
\label{R4}
\end{equation}
That is, a cooler radiation surface implies a bigger black hole. If
we further assume that the persistent emission
\,-\,$\overline{L}(0.3 - 10\,{\rm keV}) = 1.8 \times 10^{40} \,{\rm
erg \,s^{-1}}$\,-\, from M82X-2 is isotropic, we may impose a strict
Eddington limit on the mass transfer rate that can be accepted by
the black hole, $\overline{L}< L_{Edd}\simeq 1.3\times
10^{38}\,\frac{M}{M_{\odot}}\,{\rm erg \, s^{-1}}$. This imposes
stringent constraint on the lower limit of black hole mass
\begin{equation}
\begin{array}{l}
\FFr{M}{M_{\odot}}> 138.5, \quad R_{g}> 408.6\,{\rm km},
\end{array}
\label{R5}
\end{equation}
because for an accreting of $\sim 10$ per cent of the Eddington
limit - a fairly typical accretion rate for a high-state black hole
- this points towards a rough estimate of upper limit of mass $M <
1385\,M_{\odot}$.

\subsection{More accurate determination of upper mass limit of M82X-2}
For the knowledge of upper mass limit of M82X-2, a good progress can
still be made by establishing a direct physical relation between
masses of M82X-2 and M82X-1, and then one should rely on the
available mass estimates of the latter
~(e.g.~\citet{Oka,Muc,Dew,SRW,PPD,PBH,Rob,CPP,FK10,FS,PSM}). The
controversy, however, is with their mass range. As one may envisage,
a different mass estimates of M82X-1 may yield different mass values
of M82X-2. So, we should be careful in choosing the most accurate
black hole mass measurement to-date. As the centroid of the
persistent emission is between M82X-2 and M82X-1 which indicates
that M82X-2 harbors an IMBH, likewise M82X-1, we suppose that
accretion onto a black hole is well approximated by the relation
$L_{acc}=\eta c^{2}\dot{M}=\frac{GM}{R}\dot{M}$. This gives
$\dot{M}_{1}/\dot{M}=\overline{L}_{1}/\overline{L}\simeq 5.556$.
According to MTBH, we have $\dot{M}\propto M^{2}$ for both the
collisionless and the hydrodynamic spherical accretions onto black
hole~\citep{gago2,gago3}. Making use of these relations gives
\begin{equation}
\begin{array}{l}
M_{1}\simeq 2.357 M.
\end{array}
\label{R10}
\end{equation}
The M82X-1 is a good candidate for hard state ULXs which may be one
of the very few ULXs that change their spectral state during
outbursts, switching from a hard to a thermal state~\citep{FS}. The
"type-C" low frequency quasi-periodic oscillations  and broadband
timing noise, detected in the two XMM-Newton observations in 2001
and 2004, in the central region of M82~\citep{SM,Dew,Muc} and later
confirmed to originate from the M82X-1~\citep{FK07}, suggest that
the ULX harbors a massive black hole. The mass estimate
by~\citep{Dew} is based on the assumption that M82X-1 also follows
well established relation of the photon spectral index versus QPO
frequency, $\Gamma-\nu_{QPO}$, found for the Galactic X-ray binaries
in their high or intermediate states. The resulting IMBH mass for
M82X-1 is in the range of $25-520\,M_{\odot}$. However, there may be
systematic errors in the photon indices measured with XMM-Newton and
Rossi X-ray Timing Explorer (RXTE) due to contamination from nearby
sources, as indicated by large apparent changes in the effective
absorption column. Another mass estimate by~\citep{CPP}  is inferred
from the correlations with the X-ray luminosity and type C QPO
frequency. This method is based on the correlation between
characteristic frequencies, on the fundamental plane and on the
variability plane of accreting black holes. Exploring this method,
the black hole masses inferred from the characteristic frequencies
are all about $10^{3}-10^{4}\,M_{\odot}$ indicating that ULXs
contain IMBHs~\citep{CPP,FS}. But, these results were not without
problems, notably pointed out by the same authors. Such mass
estimates are based on scaling relations which use low-frequency
characteristic timescales which have large intrinsic uncertainties.
In particular, it was unclear whether these mHz oscillations are
indeed the Type-C analogs of stellar mass black, and both the Type-C
and the mHz oscillations are variable, resulting in a large
dispersion in the measured mass of $25-1300\,M_{\odot}$. Positive
identification of the emission states requires both timing and
spectral information. Consequently, with simultaneous observations
exploiting the high angular resolution of Chandra to isolate the ULX
spectrum from diffuse emission and nearby sources and the large
collecting area of XMM-Newton observations of M82 to obtain timing
information, ~\citet{FK10} found surprisingly that the previously
known QPOs in the source disappeared. The light curve was no longer
highly variable and the power spectrum was consistent with that of
white noise. The energy spectrum also changed dramatically from a
straight power-law to a disk-like spectrum. The disappearance of
QPOs and the low noise level suggest that the source was not in the
hard state. All results are well consistent with that expected for
the thermal state. The monitoring data from RXTE indicate that these
Chandra and XMM-Newton observations were made during the source
outbursts, suggesting that M82X-1 usually stays in the hard state
and could transition to the thermal state during outbursts. The
spectral fitting suggests that the ULX contains a close to Eddington
$(L_{disk}/L_{Edd} \sim 2)$ rapidly spinning IMBH of
$200-800\,M_{\odot}$ masses. The thermal dominant states are all
found during outbursts. Nonetheless, modeling of X-ray energy
spectra during the thermal-dominant state using a fully relativistic
multi-colored disk model has large uncertainties owing to both
systematic and measurement errors. In addition to the large mass
uncertainty associated with the modeling, the same authors also
found that the energy spectra can be equally well-fit with a
stellar-mass black hole accreting at a rate of roughly $160$ higher
than the Eddington limit. Also, the X-rays from this source are
known to modulate with an orbital periodicity of 62 days, which
indicates to an intermediate-mass black hole with mass in the range
of $200-5000\,M_{\odot}$~\citep{PPD,PBH}. But, a recent study finds
that this periodicity may instead be due to a precessing accretion
disk in which case a stellar-mass black hole will suffice to explain
the apparent long periodicity~\citep{PS}. Thus, the mass
measurements above have large uncertainties. This makes black hole
masses obtained by this method at the very least questionable. In
what follows, therefore, we adopt an alternative and less ambiguous
mass determination for intermediate-mass black holes suggested
by~\citep{PSM} which seems to be a more reliable determinant of the
mass of M82X-1. These authors searched RXTE's proportional counter
array archival data to look for $3:2$ oscillation pairs in the
frequency range of $1-16$ Hz which corresponds to a black hole mass
range of $50-2000\, M_{\odot}$. In stellar-mass black holes it is
known that the high frequency quasi-periodic oscillations that occur
in a $3:2$ ratio ($100-450$ Hz) are stable and scale inversely with
black hole mass with a reasonably small
dispersion~\citep{McCR,RMBO,RMMc,RSM,Stro1,Stro2}. ~\citet{PSM}
report stable, twin-peak (3:2 frequency ratio) X-ray quasi-periodic
oscillations from M82X-1 at the frequencies of $3.32\pm 0.06$ Hz
(coherence, Q = centroid frequency $(\nu)/$width$(\nu) > 27)$ and
$5.07\pm 0.06$ Hz $(Q > 40)$. The discovery of a stable $3:2$
high-frequency periodicity simultaneously with the low-frequency mHz
oscillations allows for the first time to set the overall frequency
scale of the X-ray power spectrum. This result not only asserts that
the mHz quasi-periodic oscillations of M82X-1 are the Type-C analogs
of stellar-mass black holes but also provides an independent and the
most accurate black hole mass measurement to-date. Assuming that one
can scale the stellar-mass relationship, they estimate the black
hole mass of M82X-1 to be $428\pm 105\,M_{\odot}$. They also
estimate the mass using the relativistic precession model, which
yields  a value of $415\pm 63\,M_{\odot}$. Combining the average
2-10 keV X-ray luminosity~\cite{PS,KF}  of the source of $5\times
10^{40}\,{\rm ergs\, s^{-1}}$ with the measured mass suggests that
the source is accreting close to the Eddington limit with an
accretion efficiency of $0.8\pm 0.2$.

Making use of the mass values $428\pm 105\,M_{\odot}$
with~(\ref{R10}), we provide, therefore, the mass estimate for
M82X-2:
\begin{equation}
\begin{array}{l}
M\simeq 138.5-226\,M_{\odot}, \quad R_{g}\simeq 408.6-666.7\,{\rm
km},
\end{array}
\label{R13}
\end{equation}
Rotation speed at surface of M82X-2 with upper limit mass
$226\,M_{\odot}$ as rigidly spinning IMBH configuration of angular
velocity $\Omega$ equals $v=R_{g}\Omega\simeq 3.06\times
10^{8}\,{\rm cm\,s^{-1}}$. We also have slightly improved the lower
mass limit $323\,M_{\odot}$ of M82X-1 given by~\citep{PSM} now to be
$ 326.5\,M_{\odot}$. Combining~(\ref{R13}) and ~(\ref{R4}), we
obtain then
\begin{equation}
\begin{array}{l}
2.06\times 10^{33} < \FFr{T^{4}_{s}}{K^{4}}\,\FFr{d}{\rm m} <
3.34\times 10^{33}.
\end{array}
\label{R666}
\end{equation}
For reasons that will become clear below, next, in our setting we
retain the rather concrete proposal of non-spinning black
holes~\citep{gago2,gago5}, i.e. the neutrino flux from spinning
M82X-2 might as well be highly beamed along the plane of
proto-matter disk and collimated in very small opening angle. For
the values~(\ref{R13}), this yields the constraints
\begin{equation}
\begin{array}{l}
7.5\times 10^{-7}\,\FFr{d}{\rm m}< \theta_{\nu}\sim \varepsilon_{d}<
1.2\times 10^{-6}\,\FFr{d}{\rm m}.
\end{array}
\label{R7777}
\end{equation}
Besides, the $\varepsilon_{d}$ is likely to be about an order of
magnitude $\sim 10^{-5}$ for M82X-2. Therefore,
\begin{equation}
\begin{array}{l}
d\simeq 61- 100\,{\rm m},\quad \varepsilon_{d}\simeq (4.6-7.5)\times
10^{-5},
\end{array}
\label{R777}
\end{equation}
is a good guess for the thickness of proto-matter disk at the edge
of event horizon. This, together with~(\ref{R666}), give $
T_{s}\simeq 7.6\times 10^{7}$ K. Thus, M82X-2 indeed releases $\sim
99.6\%$ of its radiative energy predominantly in the X-ray bandpass
of $0.3-30$ keV. However, the studies in other wavelengths well give
us useful information on its physical nature and environment. From
Wien's displacement law we obtain the wavelength
$\lambda_{max}\simeq 0.381$nm at which the radiation curve peaks,
which corresponds to energy $h\nu_{max}\simeq 3.2$ keV. As an
immediate corollary to the assumption that the emission arisen from
accretion is isotropic, we are able to infer the most important
ratios of the pulsed and persistent luminosities to the isotropic
Eddington limit for M82X-2:
\begin{equation}
\begin{array}{l}
\FFr{\widetilde{L}}{L^{}_{Edd}}\simeq 0.17- 0.28,\quad
\FFr{\overline{L}}{L^{}_{Edd}}\simeq 0.63-1.03,
\end{array}
\label{Ed1}
\end{equation}
respectively, where $ L_{Edd}\simeq (1.75-2.85)\times 10^{40}\,{\rm
erg\,s^{-1}}$. These properties appear consistent with the
sub-Eddington hard state, which given the observed luminosities of
M82X-2 suggests the presence of SIMBH with a dynamical
mass~(\ref{R13}). Given  the angular velocity $\Omega=\frac{2\pi}{P}
(P=1.37\,{\rm s})$, we may calculate the rotational kinetic energy
$E^{}_{rot}=\frac{1}{2}I\Omega^{2}$ of M82X-2, where
$I=\frac{2}{5}MR^{2}_{g}$ is the moment of inertia if M82X-2 is
regarded as the rigidly spinning {\em canonical} configuration of
mass $M$ and radius $R_{g}$. Hence
\begin{equation}
\begin{array}{l}
E^{}_{rot}\simeq (3.72-16.17)\times 10^{51}\,{\rm erg}.
\end{array}
\label{R16}
\end{equation}
When all energy thermalized, radiation emerges as a blackbody. A
significant fraction of the accretor M82X-2 surface radiates the
accretion luminosity at temperature
\begin{equation}
\begin{array}{l}
T_{b}=\left(\FFr{L^{}_{acc}}{4\pi R^{2}_{g}}\right)^{1/4}\simeq
\left(\FFr{L^{}_{Edd}}{4\pi R^{2}_{g}}\right)^{1/4},
\end{array}
\label{R17}
\end{equation}
such that $T_{b}\simeq (3.88-5.37)\times 10^{6}K. $ The
gravitational energy of each accreted electron-proton pair turned
directly into heat at (shock) temperature $T_{sh}$:\,
$3kT_{sh}=\frac{GMm_{p}}{R_{g}}$, so $T_{sh}\simeq 1.8\times
10^{12}K$. Hence typical photon energies of persistent radiation
lies between $kT_{b}\simeq 0.34\, {\rm keV}\leq h\nu \leq
kT_{sh}\simeq 180\, {\rm MeV}$. So, M82X-2 is persistent X-ray and
possibly gamma-ray emitter. Also, given the mass of the most
brightest source M82X-1 of persistent X-ray radiation,  typical
photon energies of persistent radiation lie in range
$kT_{b(X1)}\simeq 0.3\, {\rm keV}\leq h\nu  \leq kT^{}_{sh}\simeq
180\, {\rm MeV}$.

\subsection{The mass of companion star and orbit parameters}
Once the mass scaling of M82X-2 is accomplished, this can
potentially be used further to quantify the association between the
M82X-2 and the optical secondary donor star in X-ray binary. The
orbital period $P_{or}$ is a key parameter for dynamical mass
measurement. From Fig.~\ref{F2}, the separation of the two masses is
$a$, and their distances from the center of mass are $a_{1}$ and
$a_{2}$. The highly circular orbit, combined with the mass function
$f(M,\,M_{2},\,i)= 2.1 M_{\odot}$, the lack of eclipses and
assumption of a Roche-lobe- filling companion constrain the
inclination angle to be $i<60^{o}$~\citep{NuS}. These alow to
determine the mass $M_{2}$ of donor star:
\begin{equation}
\begin{array}{l}
\FFr{M_{2}}{M_{\odot}}> \left\{\begin{array}{l} 48.3,\quad
\mbox{for}\quad
M=138.5\,M_{\odot},  \\
64.9,\quad \mbox{for}\quad M=226\,M_{\odot}.
\end{array}
\right.
\end{array}
\label{R20}
\end{equation}
Thus, optical companion is a typical O/B supergiant, which evolves
away from the main sequence in just a few million years. The binary
separation can be computed with Kepler's Law
$a^{3}=\frac{G(M+M_{2})}{4\pi^{2}}P^{2}_{or}$.  The Doppler curve of
the spectrum of NuSTAR J095551+6940.8 shows a $P_{or}=2.5$-day
sinusoidal modulation arisen from binary orbit~\citep{NuS}. All
these give the projection of the orbital velocity of M82X-2 along
the line of sight $ v_{1}=\frac{2\pi}{P_{or}}a_{1}\sin i\simeq
200.9\frac{\rm km}{\rm s}$,  and hence $a_{1}\sin i\simeq 9.9
R_{\odot}$. The absence of eclipses implies $ \cos i> \frac{R}{a}$,
where $R$ is the radius of companion donor star. Hence
\begin{equation}
\begin{array}{l}
R>\left\{\begin{array}{l} 22.1\,R_{\odot},\quad \mbox{for}\quad
M=138.5\,M_{\odot},  \\
25.7\,R_{\odot},\quad \mbox{for}\quad M= 226\,M_{\odot}.
\end{array}
\right.
\end{array}
\label{R24}
\end{equation}
As well as $a_{1}>11.4\,R_{\odot}$, $a_{2}>32.9\,R_{\odot}$ for
$M/M_{\odot}=138.5$; and $a_{1}$ is the same,
$a_{2}>39.9\,R_{\odot}$ for $M/M_{\odot}=226$. The Roche lobe radius
$R_{L}$ for donor star is
\begin{equation}
\begin{array}{l}
R_{L}>\left\{\begin{array}{l} 15\,R_{\odot},\qquad \mbox{for}\quad
M=138.5\,M_{\odot},  \\
14.3\,R_{\odot},\quad \,\mbox{for}\quad M= 226\,M_{\odot}.
\end{array}
\right.
\end{array}
\label{R2424}
\end{equation}
Thus, the M82X-2 and donor star constitute the semi-detached binary,
accreting through Roche-lobe overflow. Donor star exceeds its Roche
lobe ($R>R_{L}$), therefore its shape is distorted because of mass
transfer from donor star through the inner Lagrange point $L_{1}$ to
the M82X-2. The accretion stream is expected to be rather narrow as
it flows through the $L_{1}$ point and into the Roche lobe of the
primary.

\subsection{Spin-up rate and the torque added to M82X-2}
All accreting pulsars show stochastic variations in their spin
frequencies and luminosity, including those displaying secular
spin-up or spin-down on long time scales, blurring the conventional
distinction between disk-fed and wind-fed binaries~\citep{BeA}.
Pulsed flux and accretion torque are strongly correlated in
outbursts of transient accreting pulsars, but uncorrelated, or even
anticorrelated, in persistent sources. The observed secular spin-up
rate can be accounted for quantitatively if one assumes the
reduction of the torque on the rapidly spinning object. Continuing
on our quest, below we determine the conditions under which pulsed
source M82X-2 spins up and hance gains rotational energy as matter
is accreted, i.e. we discuss the relationship between the properties
of the exterior flow and the measured rate of change of angular
velocity $d\Omega/dt$. Although measurements of spin-up or spin-down
appears to be the most promising method for determining the angular
momentum transport by the inflowing matter, which in turn, may
provide information about pattern of material flow outside the event
horizon of SIMBH, the extraction of this information from such
measurements clearly requires some care. We explore the relationship
between the torque ($\equiv l$) flux through the event horizon,
spin-up rate of SIMBH and the rate of change of its rotational
energy. The rates of change of the SIMBH angular velocity and of the
rotational energy can be related to the flux of torque across the
event horizon boundary as follows. The rate of change of the torque
is given by \citep{GLP}
\begin{equation}
\begin{array}{l}
\FFr{d}{dt}(I\Omega)=\dot{M}l,
\end{array}
\label{R25}
\end{equation}
where $I$ is the moment of inertia,  and $l$ is the torque added to
the SIMBH per unit mass of accreted matter. Equation ~(\ref{R25})
gives for the rate of change of angular velocity
\begin{equation}
\begin{array}{l}
\FFr{d\Omega}{dt}=\FFr{\dot{M}}{L_{bh}}\left[l\Omega-\Omega^{2}R^{2}_{gir}\left(\FFr{M}{I}\FFr{dI}{dM}\right)\right],
\end{array}
\label{R26}
\end{equation}
where $L_{bh}\equiv I\Omega$, and $R_{gir}$ is the radius of
gyration of SIMBH. The rate of change of the rotational energy is
\begin{equation}
\begin{array}{l}
\FFr{dE_{rot}}{dt}=\FFr{d}{dt}\left(\FFr{1}{2}I\Omega^{2}\right).
\end{array}
\label{R27}
\end{equation}
To make further progress we recast~(\ref{R26}) and~(\ref{R27}) into
the form
\begin{equation}
\begin{array}{l}
\FFr{dE_{rot}}{dt}=\dot{M}\left[l\Omega-\FFr{1}{2}\Omega
R^{2}_{gir}\left(\FFr{M}{I}\FFr{dI}{dM}\right)\right].
\end{array}
\label{R28}
\end{equation}
When $\frac{M}{I}\frac{dI}{dM}>0$, which is generally the case, the
SIMBH's behavior can be conveniently characterized by the
dimensionless parameter
\begin{equation}
\begin{array}{l}
\zeta\equiv \FFr{l}{\Omega
R^{2}_{gir}}\left(\FFr{M}{I}\FFr{dI}{dM}\right)^{-1}.
\end{array}
\label{R29}
\end{equation}
Thus the black hole loses rotational energy and spins down for
$\zeta<1/2$, whereas it gains rotational energy and spins up for
$\zeta>1$; for $1/2< \zeta< 1$ the black hole spins down even though
it is gaining rotational energy. The logarithmic derivative
$\frac{M}{I}\frac{dI}{dM}$ for  {\em canonical} configuration, i.e.
spinning uniform-density sphere with mass $M$ and radius $R_{g}$, is
$\frac{d\ln I}{d\ln M}=\frac{d}{d\ln M}\ln
\left(\frac{2}{5}MR^{2}_{g}\right)=3$, so
\begin{equation}
\begin{array}{l}
\zeta=\FFr{l}{3\Omega R^{2}_{gir}},
\end{array}
\label{R30}
\end{equation}
where $R^{2}_{gir}=I/M=\frac{2}{5}R^{2}_{g}$. For the spin up regime
of M82X-2 when $\zeta\simeq 0.073\frac{l}{\Omega R^{2}_{gir}}>1$, we
obtain $l> 2.192 R^{2}_{g}\,{\rm s^{-1}}$, so
\begin{equation}
\begin{array}{l}
l> \left\{\begin{array}{l} 3.7\times 10^{5}\,{\rm km^{2}
s^{-1}},\quad \mbox{for}\quad M=138.5\,M_{\odot}\\
9.7\times 10^{5}\,{\rm km^{2} s^{-1}},\quad \mbox{for}\quad M=
226\,M_{\odot}.
\end{array}
\right.
\end{array}
\label{R31}
\end{equation}
The time derivative of the angular velocity~(\ref{R26}) gives
\begin{equation}
\begin{array}{l}
l=\FFr{4\pi
R^{2}_{g}}{5P}\left(3-\FFr{\dot{P}}{P}\FFr{M}{\dot{M}}\right).
\end{array}
\label{R32}
\end{equation}
Combing $\dot{M}\simeq 6.35\times 10^{-7}\,M_{\odot}\,yr^{-1}$, and
a linear spin-up $\dot{p}$ of the NuSTAR J095551+6940.8 pulsar, from
~(\ref{R32}) we obtain the torques added to M82X-2 per unit mass of
accreted matter, which satisfy the spin-up condition~(\ref{R31}) of
$\zeta>1$:
\begin{equation}
\begin{array}{l}
l\simeq \left\{
\begin{array}{l}
1.1\times 10^{11}\,{\rm  km^{2} s^{-1}},\quad \mbox{for}\quad
M= 138.5\,M_{\odot}, \\
9.8\times 10^{11}\,{\rm km^{2} s^{-1}},\quad \mbox{for}\quad
M=226\,M_{\odot}.
\end{array}
\right.
\end{array}
\label{R33}
\end{equation}
This is not a final report on a closed subject, and needless to say
once again that astronomers are not yet at the end of producing a
coherent picture of the ULX Universe, so there is plenty to argue
about. We believe that there is still very much to be gained by
further study of the issues that we raised in this paper. There are
deep conceptual and technical problems involved, and these provide
scope for the arguments discussed, which are carefully presented in
both mathematical and physical terms. It should be emphasized that
the key to our construction procedure of both MTBH in general, and
suggested model of mighty ultraluminous X-ray pulsations from
periodic source M82X-2 in particular, is widely based on the
premises of our experience of accretion physics. Therefore, what we
have presented here has all the vices and virtues of the classical
scenario of runaway core collapse which has always been a matter of
uncertainties and controversies. On the other sight, astronomers
doubted the existence of the superdense proto-matter sources, away
from the accretion physics. Models of this type have a long
tradition of precursors that dates back to the pioneering seminal
papers by \citet{b2}. Since then the very possibility of existence
of such sources was rejected, and the idea of the superdense
proto-matter was dismissed as a fancy of eccentric astronomer.
However, we caution that our entire constructions will be valid as
well in the case if some hitherto unknown yet mechanism in Nature
will in somehow or other way produce the superdense proto-matter,
away from the accretion physics.

\section{Concluding remarks}
Our analysis of periodic source M82X-2 goes quite on the contrary to
the conventional wisdom. The pulsed luminosity of M82X-2 is the most
extreme violation of the Eddington limit and could be reconciled
with that in the model of common pulsar only by very arbitrary
assumptions on geometric beaming of accretion flow on neutron star.
Instead of making such assumptions, our approach finds it preferable
to return in part to the ideas of MTBH framework to circumvent the
alluded obstacles without the need for significant breaking of
Eddington limit. The new conceptual element of the implications of
MTBH in tackling this problem is noteworthy. This explores the most
important processes of spontaneous breaking of gravitation gauge
symmetry and rearrangement of vacuum state at huge energies, making
room for growth and merging behavior of black holes. Putting apart
the discussion of inherent problems of the mass scaling of the black
holes in ULXs, we have focused on black hole. We assume the M82X-2
is being SIMBH, resided in the final stage of growth. If this were
the case, eventually, a thin co-spinning proto-matter disk driven by
accretion would be formed around the spinning proto-matter core. It
was tilted from the plane of accretion on a definite angle $\delta$
towards the equator and has reached out the edge of the event
horizon, where a metric singularity inevitably disappears. The
energy is carried away then from event horizon through a thin belt
area to outside world by both the ultra-high energy neutrinos
produced in the superdense proto-matter medium, and the thermal
defuse blackbody radiation released from the outer surface layers of
ordinary matter of spinning SPC and co-spinning proto-matter disk.
All of the rotational energy of SIMBH is dissipated as thermal
defuse blackbody X-ray radiation to outside world. As M82X-2 spins,
we see pulses because of the axial tilt or obliquity. That is, the
M82X-2 is the emitter of both persistent and pulsating X-rays. We
derive the general profiles of pulsed luminosity and X-ray flux of
M82X-2. Hence M82X-2 indeed releases $\sim 99.6\%$ of its radiative
energy predominantly in the X-ray bandpass of $0.3-30$ keV. The
resulting theoretical model necessarily includes a number of poorly
known parameters. We give a quantitative account of all the
energetics, a dynamical mass scaling and orbital parameters of the
semi-detached X-ray binary containing primary M82X-2 and the
secondary massive O/B-type donor star, accreting through Roche-lobe
overflow. The position angles can be evaluated from rigorous
comparison with the behavior of observed pulsed light curve of
M82X-2, which will be discussed in a future publication.

\section*{Acknowledgments}
The very helpful and knowledgable comments and positive feedback
from the anonymous referee that have essentially improve the quality
of the final manuscript are much appreciated. I would like to thank
H.Pikichyan and A.Beglaryan for useful discussions.

\nocite{*}
\bibliographystyle{spr-mp-nameyear-cnd}
%\bibliography{myref}
\bibliography{biblio-u1}

\end{document}